\documentclass[epsf,useAMS,usenatbib, usegraphicx]{mn2e}
\usepackage{graphicx}
\usepackage{epsfig}
\usepackage{amsmath}
\usepackage{amssymb}
\usepackage{rotating}
\usepackage{color}
\usepackage{pifont}
\usepackage{longtable}
\usepackage{pdflscape,lscape}
\usepackage{float}
\usepackage{placeins}
\usepackage{pdfpages}
\usepackage[T1]{fontenc}
\usepackage{bold-extra}

\newcommand{\ion}[2]{{#1}\,{\sc #2}}
\newcommand{\teff}{$T_{\rm eff}$}
\newcommand{\logg}{$\log g$}
\newcommand{\vsini}{$v \sin i$}
\newcommand{\kms}{km\,s$^{-1}$}
\newcommand{\dst}{$\delta$\,Sct}
\newcommand{\gd}{$\gamma$\,Dor}
\newcommand{\cd}{d$^{-1}$}

\usepackage{appendix}
\usepackage{booktabs}
\usepackage{upgreek}
\usepackage{url}

\usepackage{natbib}
\title[Hot $\gamma$ Doradus and A-F hybrid stars]{Spectroscopy of hot $\gamma$ Doradus and A-F hybrid {\it Kepler} candidates close to the hot border
of the $\delta$\,Scuti instability strip}

\author[F. Kahraman Ali\c{c}avu\c{s} et al.]{F. Kahraman Ali\c{c}avu\c{s}$^{1,2}$\thanks{E-mail: filizkahraman01@gmail.com},
E. Poretti$^{3,4}$\thanks{E-mail: ennio.poretti@tng.iac.es}, G. Catanzaro$^{5}$, B. Smalley$^{6}$, E. Niemczura$^{7}$,\and M. Rainer$^{3,8}$, G. Handler$^{1}$
\\
$^{1}$Nicolaus Copernicus Astronomical Center, Bartycka 18, PL-00-716 Warsaw, Poland\\
$^{2}$\c{C}anakkale Onsekiz Mart University, Faculty of Sciences and Arts, Physics Department, 17100, \c{C}anakkale, Turkey\\ 
$^{3}$INAF - Osservatorio Astronomico di Brera, via E. Bianchi 46, 23807 Merate (LC), Italy\\
$^{4}$INAF - Fundación Galileo Galilei, Rambla Jos\'{e} Ana Fernandez P\'{e}rez 7, 38712 Bre\~{n}a Baja, TF, Spain\\
$^{5}$INAF - Osservatorio Astrofisico di Catania, Via S. Sofia 78, I-95123 Catania, Italy\\
$^{6}$Astrophysics Group, Keele University, Staffordshire, ST5 5BG, UK \\
$^{7}$Instytut Astronomiczny, Uniwersytet Wroc\l{}awski, ul. Kopernika 11, 51-622 Wroc\l{}aw, Poland\\
$^{8}$INAF - Osservatorio Astrofisico di Arcetri Largo E. Fermi 5, 50125 Firenze, Italy\\
}

\begin{document}

\date{Accepted ... Received ...; in original form ...}

\pagerange{\pageref{firstpage}--\pageref{lastpage}} \pubyear{2019}

\maketitle

\label{firstpage}

\begin{abstract}

If \gd-type pulsations are driven by the convective blocking mechanism, a convective envelope at a sufficient depth is essential. 
There are several hot \gd\, and hybrid star candidates in which there should not be an adequate convective 
envelope to excite the \gd-type oscillations. The existence of these hot objects needs an explanation. 
Therefore, we selected, observed and studied 24 hot \gd\, and hybrid candidates to investigate their properties.
The atmospheric parameters, chemical abundances and \vsini\, values of the candidates were obtained using  
medium-resolution ($R$\,=\,46\,000) spectra taken with the FIES instrument mounted at the Nordic Optical Telescope. 
%and binarity effects were determined in those hot objects. 
We also carried out frequency analyses of the \textit{Kepler} long- and short-cadence data to determine the exact pulsation contents. 
We found only five bona-fide hot \gd\, and three bona-fide hot hybrid stars in our sample. The other 16 stars were found to be 
normal \gd, \dst, or hybrid variables. 
% thus suggesting that the initial sample is largely (70\%) contaminated. 
No chemical peculiarity was  detected in the spectra of the bona-fide hot \gd\, and hybrid stars.
We investigated the interplay between rotation  and pulsational modes.
%and in particular the hypothesis that hot \gd\, stars could be actually rapidly-rotating slowly-pulsating B variables, 
%that In particular, 
We also found that the hot \gd\, stars have higher {\it Gaia} luminosities and larger radii compared to main-sequence A-F stars.
%thus supporting the link with B-type stars. 
%The higher {\it Gaia} luminosities may also support the binary nature of these stars which could have a far luminous companion with a long 
%orbital period.

\end{abstract}

\begin{keywords}
stars: general -- stars: abundances -- stars: atmospheres -- stars: rotation -- stars: variables: $\gamma$ Doradus
\end{keywords}

\section{Introduction} \label{intro}

The {\it Kepler} spacecraft \citep{2010ApJ...713L..79K, 2010Sci...327..977B} was originally launched to detect Earth-like transiting planets, but 
it also shed new light on many other aspects of stellar astrophysics \citep[e.g.][]{2019arXiv190605587P, 2019MNRAS.485..560C, 2019MNRAS.484..451H}.
%. However, many new variable stars have been discovered thanks to its high precision photometry. Especially,
In particular, new discoveries about pulsating stars
provided important input for asteroseismology, opening new frontiers and posing new questions.
Some of those concern the classes of the A- and F-type
pulsating variables located on or close to the main sequence, i.e., 
$\delta$\,Scuti (\dst) and $\gamma$\,Doradus (\gd) stars.

The \dst\, stars generally exhibit pressure ($p$) %, gravity $g$ and mixed ($p$ and $g$) 
modes which are excited by the $\kappa$ mechanism in the \ion{He}{ii} ionization zone. 
They typically oscillate with frequencies higher than $\sim$5~d$^{-1}$. 
The \dst\, stars are mostly placed in the lower part of the classical instability strip. 
Recently, \citet{2019MNRAS.485.2380M} classified a sample of over 15\,000 {\it Kepler} A-type  and F-type targets into
\dst\, and non-\dst\, stars, also providing a subdivision in groups on the basis of the observed pulsation properties. They also
defined a new empirical instability strip for \dst\, stars.

The \gd\, stars are a little cooler than the \dst\, variables. The gravity ($g$) mode oscillations of 
\gd\, stars are believed to be excited by the convective blocking mechanism \citep{2000ApJ...542L..57G} with frequencies typically 
lower than 5~\cd. 
In a recent study, it was suggested that the \gd-type pulsations are caused by the combination of $\kappa$ mechanism and 
the convection-oscillation coupling \citep{2016MNRAS.457.3163X}. 
%These variables are also located the lower part of the classical instability strip.
The \gd\, domain
partially overlaps the cool border of the theoretical instability strip of \dst\, stars. In this part, new variables 
called A-F type hybrid stars were predicted \citep{1999MNRAS.309L..19H, 2004A&A...414L..17D}. These stars display both \dst\, and \gd\, type pulsations,
i.e., $p$ and $g$ modes. 

Before {\it Kepler}, only a few
hybrid stars had been discovered by means of ground-based observations \citep{2005AJ....129.2026H,2008A&A...489.1213U,2009MNRAS.398.1339H} and
their positions in the Hertzsprung-Russell (H-R) diagram  matched very well with %were found inside the 
the theoretically predicted region. The precision and near continuous nature of {\it Kepler} photometry revolutionized the field, showing that apparently there are many
A-F type hybrid stars located beyond the theoretical region \citep{2010ApJ...713L.192G, 2011A&A...534A.125U}.
On the other hand, it was also found that some \dst\, and \gd\, stars are also placed outside the respective instability strips.
These results give rise to conflicts with theory. Therefore, it was timely to fix the exact positions
of these variables  in the H-R diagram and to revise the borders of their instability strips. For these reasons, spectroscopic
studies were carried out to determine accurate atmospheric parameters of \dst, \gd, and hybrid
stars (e.g. \citealt{2012MNRAS.422.2960T, 2015MNRAS.450.2764N, 2017MNRAS.470.2870N}).
These investigations  confirmed that some stars are actually outside the theoretical instability strips.

In particular, \citet{2014MNRAS.437.1476B} and \citet{2016MNRAS.460.1318B} 
noticed how 
 some gravity-mode pulsators are located close to, and even beyond,   
 the hot border of \dst\, instability strip, defining them as hot \gd\, stars. These variables are 
remarkable objects because 
if the theory of convective driving would apply to stars 
located in this part of the H-R diagram then they should not have a  
sufficient convective envelope to drive the \gd-type pulsations. 
It has been suggested that the hot \gd\, stars could be A- or B-type stars with a cooler
\gd\, companion,  or simply stars with wrong determinations of the effective temperature (\teff).
A spectroscopic study of hot \gd\, stars was undertaken \citep{2016MNRAS.460.1318B} and
%The hot \gd\, stars were discussed by \citet{2016MNRAS.460.1318B},
%who suggested that they  could be A- or B-type stars with a cooler 
%\gd\, companion,  or simply stars with wrong determinations of the effective temperature (\teff). 
% Another possibility is that the low-frequency peaks are due to the rotational modulation of asymmetric intensity distributions
% on the stellar atmospheres or surfaces \citep{2015EPJWC.10106043L}.
%A spectroscopic study of hot \gd\, stars was presented by \citet{2016MNRAS.460.1318B} to derive accurate 
%\teff\, values. 
it turned out that the resulting \teff\, values were mostly consistent with those given in the 
{\sl Kepler} input catalog \citep[KIC; ][]{2011AJ....142..112B}. However, the binary nature of the stars 
was not assessed and also the chemical composition of these variables could not be probed due to the low resolution 
of the spectra.
 
Another explanation is that the hot \gd\, stars are actually rapidly-rotating slowly pulsating B (SPB) stars 
\citep{2016MNRAS.460.1318B}. 
Due to gravity darkening, their equatorial zones appear cooler than the rest of the surface, then they are classified as hot A-stars and, hence, 
as hot \gd\, variables.
If this hypothesis is true, all hot \gd\, stars should rotate with high rotational velocities. 
There are also some hot hybrid stars which are located in the same area of the H-R diagram. 
In both cases the \gd-type pulsation conflicts with the theory, 
since would require that the convective blocking mechanism is continuing to be active in the hottest A stars \citep{2014MNRAS.437.1476B}. 
The $g$ mode pulsation in those \gd\, stars can also be explained by the radiative $\kappa$ mechanism and the 
coupling between oscillation and convection \citep{2016MNRAS.457.3163X}.
%The low frequencies in these hybrid stars were tried to explain with possible companions of the stars. 
%The effects of the companions can be the reason of low frequencies. 
%{\bf This statement is unclear]} Another expşemnem ferayhlanation is that the atmospheric intensity
%distribution induced by chemical anomalies and high rotation \citep{2015EPJWC.10106043L}.
%These hot hybrid star candidates are needed an explanation as well.

\begin{figure}
\includegraphics[width=8cm,angle=0]{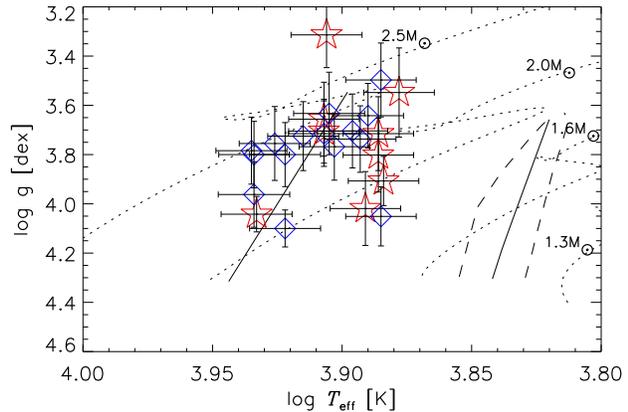}
\caption{The positions of the selected candidate $\gamma$ Dor (star symbols) and hybrid (diamonds) stars.
The parameters of the stars were taken from \citet{2014ApJS..211....2H}. The theoretical instability strips of the 
$\gamma$ Dor (dashed lines) and $\delta$ Sct (solid lines) stars were taken from \citet{2005A&A...435..927D}. 
The evolutionary tracks (Z\,=\,0.02) were 
adopted from \citet{2016MNRAS.458.2307K}.}
\label{NOT_firts_positions}
\end{figure}

In this study, a new spectroscopic survey of a selected sample of hot \gd\, and hybrid stars is presented. We focused on the candidate \gd\,
and hybrid stars located close to the hot border of the \dst\, instability strip.
We aim to answer the following questions. First, are the \teff\, values of these hot objects correct? Second,
do the hot \gd\, stars have hotter A- or B-type companions? Third, are the hybrid stars members of binary or
multiple systems? Can those hot variables be rapidly-rotating stars?
Additionally, are  hot \gd\, and hybrid stars chemically peculiar?
Therefore, we present a detailed spectroscopic analysis to determine the atmospheric parameters
(\teff, surface gravity \logg, and microturbulence velocity $\xi$), projected rotatinal velocities, binarity effects, and surface chemical abundances.

The target selection is described in Sect.~\ref{tarsel}. 
The spectroscopic observations, data reduction and normalisation in Sect.~\ref{spobs}. The determinations of the atmospheric parameters from photometric indices, spectral energy 
distribution (SED), the analysis of Balmer and metal lines are discussed in Sect.~\ref{atpar}. The chemical abundance analysis is presented in Sect.~\ref{anche}. The frequency analysis of the targets is performed in Sect.~\ref{freana}.
The discussion and conclusions are given in Sects.~\ref{disc} and \ref{conclu}, respectively.

\section{Target Selection}\label{tarsel}

We started the selection process from the  \gd\, and hybrid star candidates listed in the study 
of \citet{2011A&A...534A.125U}. They were proposed on the basis of their {\sl Kepler} light curves and preliminary  atmospheric parameters listed in the KIC. 
To better characterize them, 
we used the improved parameters reported in the revised 
{\sl Kepler} catalog of 
\citet[][hereafter H14]{2014ApJS..211....2H}, the latest available at the time of our observations (October 2016). 
This catalog is based on a compilation of literature values for atmospheric properties  derived from 
a variety of observational techniques.
%\ep{However, only %we considered that the 

High-resolution spectroscopy of a well-constrained sample of stars should provide reliable answers to the questions posed in 
%We believed that high-resolution spectroscopy of a selected sample of stars 
%constituted the best effort for providing could provide more reliable answers to the questions posed in 
Sect.\,\ref{intro}, especially when combining this technique 
with a homogeneous reduction and analysis of the spectra taken with a single instrument.
%could consitute the best effort for reliably answering
%the questions posed in Sect.~\ref{intro} could be reliably answered by 

%In the target selection process, first we selected the \gd\, and hybrid star candidates from the study of \citet{2011A&A...534A.125U}. 
%Then, we excluded the stars which have atmospheric parameters determined from high-resolution spectroscopy. 
%The atmospheric parameters (\teff\, and \logg) for the residual stars were taken from the catalog of \citet[][hereafter H14]{2014ApJS..211....2H} which gives the revised stellar parameters for {\sl Kepler} stars. 
%This catalog was used in the selection process because our observations were scheduled to be done in October 2016.} 
%
To refine our selection, we considered that the \gd\, stars have \teff\, values in the  range  6900\,$-$\,7300\,K \citep{2015ApJS..218...27V} and that 
the hybrid stars are expected to be found at the intersection  of the \gd\, and \dst\, instability strips,
 where \teff\, changes approximately from 6600 to 7300\,K. 
%The hot \gd\, stars are defined as low-frequency pulsating variables which are placed between the red border of 
%slowly pulsating B stars and the blue edge of \gd\, instability strip \citep{2014MNRAS.437.1476B, 2016MNRAS.460.1318B}. 
Given that the typical \teff\, uncertainty is $\sim$300~K \citep[][]{2017MNRAS.470.2870N} and hot \gd\, stars 
definition by \citet{2016MNRAS.460.1318B}, we selected 
 \gd\, and hybrid candidates with \teff\,$\geq$\,7500\,K. 
%taking into account a typical error of $\sim$300\,K \citep[][]{2017MNRAS.470.2870N}.}

In the end, our selection contained nine hot \gd\, and fifteen hot hybrid candidate stars 
(Table\,\ref{infotable}), for which high-resolution spectroscopy was not available. 
Some of them were observed with  low-resolution spectroscopy 
\citep{2016A&A...594A..39F, 2016MNRAS.460.1318B}, but  
it is not possible to obtain reliable \logg\, and metallicity values from this technique. 
%We considered that high-resolution spectroscopy applied to a selected sample by using the same data reduction
%and analysis can supply reliable answers to the question posed in Sect.~\ref{intro}.}
%When these targets were chosen, they did not have a spectroscopic study in the literature. However, now 

%\ep{We undertaken this project  since we considered  the 
%high-resolution spectroscopy of a selected sample of stars 
%constituted the best effort for providing reliable answers to 
%the questions posed in Sect.~\ref{intro}, especially when this technique was 
%combined with a homogeneous reduction and analysis of the spectra.
%%the high-resolution spectroscopy is still needed. Therefore, we keep these stars in our target list.} 

% We firstly considered the
% hot \gd\, and hybrid candidates in the study of \citet{2011A&A...534A.125U}. 
% We focused on the stars located close to the hot border of the \dst\, instability strip. We selected the 
% \gd\, and hybrid stars having \teff\,values in the range of 7500\,$\leqslant$\,\teff\,$\leqslant$\,8600\,K  
% in the catalog of \citet[][hereafter  H14]{2014ApJS..211....2H}. This catalog was used because our observations were scheduled to be done in October 2016. 
% As a result, nine hot \gd\, and fifteen hot hybrid candidate stars were chosen (Table\,\ref{infotable}). 
%The list of the selected stars and their general information are given in Table\,\ref{infotable}. 

The positions of the targets in the log\,\teff\,$-$\logg\, diagram are shown in Fig.\,\ref{NOT_firts_positions}. As can be 
seen from the figure, all stars are located beyond the blue edge of \gd\, instability strip. 
The \teff\, values of the selected targets were also checked by using the updated parameters of the stars \citep[][hereafter M17]{2017ApJS..229...30M}. 
It turned out that  all targets except KIC\,11508397 ($\sim$400\,K cooler) still have \teff\, values in the same range of the selection criteria. 
We used the M17 parameters in the following investigations. 

%We still kept KIC\,11508397 (\teff\,=\,7287\,K) inside our target considering a possible error in \teff.
% The H14 and M17 \teff\, values mostly agree with the spectroscopic ones within $\sim$300\,K for A-F stars \citep[][]{2017MNRAS.470.2870N, 2019MNRAS.485.2380M}. 
% Therefore, when we consider this difference, most of our targets are still hotter than the \gd\, \teff\, range ($\sim$$6900-7300$\,K, \citeauthor{2015ApJS..218...27V} \citeyear{2015ApJS..218...27V}).

% \begin{figure}
% \includegraphics[width=8cm,angle=0]{NOT_firts_positions_new.ps}
% \caption{The positions of the selected candidate $\gamma$ Dor (star symbols) and hybrid (diamonds) stars.
% The parameters of the stars were taken from \citet{2017ApJS..229...30M}. The theoretical instability strips of the 
% $\gamma$ Dor (dashed lines) and $\delta$ Sct (solid lines) stars were taken from \citet{2005A&A...435..927D}. The evolutionary tracks (Z\,=\,0.02) were 
% adopted from \citet{2016MNRAS.458.2307K}.}
% \label{NOT_firts_positions_mathur}
% \end{figure}

\section[]{Spectroscopic Observations}\label{spobs}

The stars were observed with the Fibre-fed \'{E}chelle Spectrograph (FIES),  a cross-dispersed 
spectrograph mounted on the 2.56-m Nordic Optical Telescope of the Roque de los Muchachos Observatory in La Palma \citep{2014AN....335...41T}. 
The spectrograph offers three resolving power options. The maximum resolving power is $R$\,=\,67\,000, while the medium and 
low resolving powers are $R$\,=\,46\,000 and $R$\,=\,25\,000, respectively, covering the wavelength range of $3700-8300$\,{\AA}. 

%Our targets have 10.5 V-band magnitude on average. 
Since we aimed to derive atmospheric parameters and chemical abundances of the targets, 
%and considering the average magnitude value ($V$=10.5), 
we opted for the medium-resolution ($R$\,=\,46\,000) configuration,  
taking into account the average brightness of the sample ($V$\,=\,10.6 mag).
Observations were performed in the  first halves of the nights from  October 13 to 19, 2016.
To examine the binarity nature of the targets, we tried to take at least two spectra per star on different nights. 
For some stars, only one spectrum could be taken due to the weather conditions and the limited observing time. 
%We should point out that because of the limited observation time, unfortunately the long orbital 
%period binary systems could not be detected. 
The number of the spectra for each star and the signal-to-noise (S/N) ratios at around 5500\,{\AA} for combined spectra
are given in Table\,\ref{infotable}.

The spectra were reduced by the dedicated pipeline FIEStool \citep{2014AN....335...41T}. The standard reduction procedure 
was applied. Bias subtraction, correction of flat-field, scattered light extraction, wavelength calibration, 
and merging of orders were performed for each spectrum. Normalisation of the reduced spectra was carried out 
manually by using the \textit{continuum} task of the NOAO/IRAF package\footnote{http://iraf.noao.edu/}. 
% The normalisation level of each 
% star was checked by comparing the theoretical spectra with the observed ones. The theoretical spectra were generated taking into account  
% the atmospheric parameter values given by M17.

\begin{table}
\centering
  \caption{Log of the observations (from October 13 to 19, 2016):
number of the star (N$_i$) used in the paper, KIC ID, V-magnitude, signal-to-noise ratio (S/N), 
number of taken spectra (N$_S$), and input classification \citep {2011A&A...534A.125U}.} 
    \label{infotable}
  \small
  \begin{tabular*}{0.95\linewidth}{@{\extracolsep{\fill}}lrcccr}
%   \begin{tabular}{\textwidth}{llrcccr}
  \hline
\noalign{\smallskip}
%  Number  &KIC  	  & V     &  S/N  & Number of & \multicolumn{1}{r}{Input} \\
%          &           & (mag) &       & spectra   &\multicolumn{1}{r}{classification}\\
N$_i$  &KIC  	  & V     &  S/N  & N$_s$  & \multicolumn{1}{c}{Input} \\
   &           & [mag] &       &    &\multicolumn{1}{c}{classifcation}\\
\noalign{\smallskip}
 \hline
\noalign{\smallskip}
1	      &2168333  & 10.02 &   90 & 2 & Hybrid\\
2         &3119604  & 10.90 &   50 & 1 & Hybrid\\    
3         &3231406  & 10.38 &   85 & 2 & Hybrid\\
4         &3240556  & 10.10 &   75 & 2 & Hybrid\\
5         &3245420  & 10.52 &   75 & 2 & Hybrid\\
6         &3868032  & 10.44 &   70 & 2 & \gd\\
7         &4677684  & 10.19 &   85 & 2 & \gd\\
8         &4768677  & 10.91 &   60 & 2 & Hybrid\\    
9         &5180796  & 10.11 &   85 & 2 & \gd\\
10        &5630362  & 10.69 &   70 & 2 & \gd\\
11        &6199731  & 10.94 &   50 & 1 & Hybrid\\
12        &6500578  & 10.77 &   80 & 2 & \gd\\
13        &6776331  & 10.71 &   50 & 1 & Hybrid\\
14        &7694191  & 10.94 &   45 & 1 & \gd\\    
15        &7732458  & 10.85 &   75 & 2 & Hybrid\\
16        &9052363  & 10.64 &   60 & 2 & Hybrid\\
17        &9775385  & 11.05 &   45 & 1 & Hybrid\\
18        &10281360 & 11.06 &   50 & 1 & \gd\\
19        &11197934 & 10.81 &   60 & 2 & Hybrid\\
20        &11199412 & 10.90 &   50 & 1 & \gd\\    
21        &11508397 & 10.65 &   80 & 2 & Hybrid\\
22        &11612274 & 10.44 &   85 & 4 & \gd\\
23        &11718839 & 10.73 &   60 & 2 & Hybrid\\
24        &11822666 & 10.69 &   70 & 2 & Hybrid\\
\hline
\end{tabular*}
% \begin{description}
%   \item[ ]*Only the literature spectral types of KIC\,4677684 and KIC\,9052363 are given in SIMBAD data base to be A and A2, respectively. 
%   \end{description}
\end{table}

\section{Determination of the Atmospheric parameters}\label{atpar}

%\ep{The spectroscopic atmospheric parameters (\teff, \logg, and $\xi$) were determined using the Balmer and iron lines. 
%The analysis of individual lines or narrow spectral parts ($\sim$0.5\,$-$\,1 nm) allows for
%a more reliable determination of the atmospheric parameters than the global analysis of the entire spectrum.
%This methodology has been previously used in several studies \citep[e.g., ][]{2017MNRAS.470.2870N, 2012MNRAS.422.2960T, 2010EAS....43..167N}.} 

%To obtain more reliable atmospheric parameters, we analysed individual lines or very narrow spectral parts ($\sim$0.5\,$-$\,1 nm), 
%instead of analysing entire spectrum.}

The spectroscopic atmospheric parameters (\teff, \logg, and $\xi$) were determined using the Balmer and iron lines, 
%The analysis of individual lines or narrow spectral parts ($\sim$0.5\,$-$\,1 nm) allows for
%a more reliable determination of the atmospheric parameters than the global analysis of the entire spectrum.
%This methodology has been previously used 
as done in previous several studies \citep[e.g., ][]{2010EAS....43..167N, 2012MNRAS.422.2960T, 2017MNRAS.470.2870N}. 
The approach we used in the analysis of the Balmer lines has been successfully applied in other papers \citep{2011MNRAS.411.1167C,2012MNRAS.421.1222C, 2013MNRAS.431.3258C}.
In practice, the procedure minimized the difference between observed and synthetic spectra, using the $\chi^2$ as
goodness-of-fit parameter. 
Since the rotational velocity affects the profile of the lines, we determined initial estimates of
projected rotational velocity (\vsini) values  by using the least-squares deconvolution (LSD) technique \citep{1997MNRAS.291..658D}.
The \vsini\, values were measured from the zero positions of the Fourier transform of the mean line profiles (Table\,\ref{initialvsini}).
%The initial  projected rotational velocity (\vsini) values of the stars were 
%determined to be an input value, as \vsini\, affects the profile of the balmer lines.} 
%These initial \vsini\, values were derived by using the least-squares deconvolution (LSD) technique \citep{1997MNRAS.291..658D}
%to compute the mean line profiles of the spectra. The \vsini\, values were estimated from the zero positions of the Fourier
%transform. 
%These initial values were given in Table\,\ref{initialvsini}.} 
%Additionally, using this technique the binarity nature of the targets was investigated:  only the star KIC\,3868032 was found to be 
%a single-lined spectroscopic binary. 
%The approach used in the analysis of the Balmer lines has been succesfully applied in other papers \citep{2011MNRAS.411.1167C,2012MNRAS.421.1222C, 2013MNRAS.431.3258C}.
%In practice, the procedure minimized the difference between observed and synthetic spectra, using the $\chi^2$ as
%goodness-of-fit parameter.

\begin{table*}
\centering
\footnotesize
  \caption{The atmospheric parameters taken from M17 and our \teff\,values derived from the analysis of 
  Balmer lines. ${\it E(B-V)}$ values (Sect.~\ref{gaia}) are also listed.}
  \label{initialvalues}
  \begin{tabular*}{0.95\linewidth}{@{\extracolsep{\fill}}lrccccc@{}}
\toprule 
  Number   & \multicolumn{1}{c}{KIC} &  {$E(B-V)$}     & \teff$_{M17}$         & \logg$_{M17}$          &  \teff$_{\rm Hlines}$           \\
           &                         & [mag]           &     [K]               &    [dex]               &    [K]                       \\
           &                         &  $\pm$\,0.02    &                       &                        &                 		                  \\
\midrule
   1        &2168333 &  0.03           & 8363$^{+197}_{-395}$ &  3.80$^{+0.36}_{-0.15}$ &   8000\,$\pm$\,300       \\
   2        &3119604 &  0.03           & 8383$^{+233}_{-350}$ &  4.10$^{+0.14}_{-0.15}$ &   7900\,$\pm$\,360      \\
   3        &3231406 &  0.02           & 8005$^{+222}_{-333}$ &  3.77$^{+0.40}_{-0.07}$ &   7600\,$\pm$\,270       \\
   4        &3240556 &  0.03           & 8615$^{+238}_{-374}$ &  3.78$^{+0.41}_{-0.14}$ &   8000\,$\pm$\,300      \\
   5        &3245420 &  0.11           & 8079$^{+225}_{-338}$ &  3.72$^{+0.43}_{-0.11}$ &   7900\,$\pm$\,320       \\
   6        &3868032 &  0.10           & 8564$^{+234}_{-402}$ &  4.04$^{+0.16}_{-0.16}$ &   8300\,$\pm$\,610       \\
   7        &4677684 &  0.22           & 8602$^{+68}_{-94}$   &  3.78$^{+0.28}_{-0.11}$ &   8600\,$\pm$\,480       \\
   8        &4768677 &  0.20           & 8584$^{+77}_{-86}$   &  3.76$^{+0.26}_{-0.03}$ &   8800\,$\pm$\,830       \\
   9        &5180796 &  0.15           & 8076$^{+64}_{-96}$   &  3.82$^{+0.22}_{-0.09}$ &   8100\,$\pm$\,280       \\
% 5219533 &                 & 7409         &  7705          & 7440        &               &  7700$_{+150}^{-186}$      & 242$\pm$\,   \\
    10      &5630362 & 0.11            & 7821$^{+78}_{-78}$   &  3.88$^{+0.17}_{-0.07}$ &   7700\,$\pm$\,170      \\
    11      &6199731 & 0.05            & 8040$^{+251}_{-306}$ &  3.64$^{+0.55}_{-0.09}$ &   7500\,$\pm$\,320       \\
    12      &6500578 & 0.28            & 8072$^{+223}_{-363}$ &  3.72$^{+0.42}_{-0.10}$ &   8200\,$\pm$\,450      \\
    13      &6776331 & 0.08            & 7870$^{+244}_{-325}$ &  3.70$^{+0.46}_{-0.08}$ &   7600\,$\pm$\,400       \\
    14      &7694191 & 0.19            & 8070$^{+251}_{-334}$ &  3.66$^{+0.50}_{-0.09}$ &   8300\,$\pm$\,580      \\
    15      &7732458 & 0.02            & 7766$^{+216}_{-324}$ &  3.64$^{+0.49}_{-0.09}$ &   7300\,$\pm$\,230      \\
    16      &9052363 & 0.11            & 7810$^{+216}_{-324}$ &  3.74$^{+0.42}_{-0.10}$ &   7800\,$\pm$\,360      \\
    17      &9775385 & 0.05            & 7675$^{+211}_{-316}$ &  4.05$^{+0.17}_{-0.15}$ &   7300\,$\pm$\,300      \\
    18      &10281360& 0.06	       & 7776$^{+216}_{-325}$ &  4.02$^{+0.22}_{-0.14}$ &   7200\,$\pm$\,280      \\
    19      &11197934& 0.06            & 7641$^{+68}_{-91}$   &  3.83$^{+0.22}_{-0.06}$ &   7600\,$\pm$\,360      \\
    20      &11199412& 0.02            & 7693$^{+239}_{-319}$ &  3.72$^{+0.46}_{-0.08}$ &   7200\,$\pm$\,280      \\
    21      &11508397& 0.00            & 7287$^{+76}_{-87}$   &  3.87$^{+0.18}_{-0.10}$ &   7200\,$\pm$\,240       \\
    22      &11612274& 0.00            & 7694$^{+214}_{-322}$ &  3.60$^{+0.54}_{-0.06}$ &   7000\,$\pm$\,190      \\
    23      &11718839& 0.03            & 8443$^{+233}_{-367}$ &  3.76$^{+0.42}_{-0.14}$ &   8100\,$\pm$\,500      \\
    24      &11822666& 0.03            & 8615$^{+238}_{-374}$ &  3.78$^{+0.42}_{-0.14}$ &   8200\,$\pm$\,540      \\
\bottomrule
\end{tabular*}
\end{table*} %$E(B-V)$

% defined as

% \begin{equation}
% \chi^2\,=\,\frac{1}{N}\sum \left(\frac{I_{obs} - I_{th}}{\delta I_{obs}}\right)Response: The questions have been answered in the sect. 5.2.
% \end{equation}

% \noindent
% where N is the total number of points, I$_{obs}$ and I$_{th}$ are the intensities of the observed and computed profiles, respectively,
% and $\delta I_{obs}$ is the photon noise. 
Synthetic spectra were generated in three steps. First, we computed LTE atmospheric models
using the ATLAS9 code \citep{1993KurCD..13.....K,kur93}. Second, the stellar spectra were then synthesized using SYNTHE \citep{1981SAOSR.391.....K}. 
Third, the spectra were convolved with the instrumental and rotational broadenings.

\teff\, was estimated by computing the ATLAS9 model atmosphere which gave the best match
between the observed $H_{\alpha}$, $H_{\beta}$, $H_{\gamma}$, and $H_{\delta}$ lines profile and those computed with SYNTHE.
The models were computed using solar opacity distribution functions (ODF). It is also known 
that Balmer lines are not sensitive to \logg\, parameter for \teff\,$\lesssim$ 8000\,K \citep{2002A&A...395..601S}. 
Considering the \teff\, range of our targets and the error bars, we fixed the \logg\, and metallicity values to
4.0\,dex and solar, respectively.

The Balmer lines are located far from the edges of the \'{e}chelle orders.
% so that it was possible to safely recover the whole profiles. 
The simultaneous fitting of four lines led to a final solution
at the intersection of the four $\chi^2$ iso-surfaces. An important source of uncertainty arose from 
difficulties in the normalization since it is always challenging for Balmer lines in \'{e}chelle spectra. 
The uncertainties on the \teff\, values were estimated by introducing a 1$\sigma$ change in the 
normalization level and a 0.2\,dex error on \logg\, and metallicity. 
We also considered the  errors on  the initial \vsini\, values (Table~\ref{initialvsini}).
%The uncertainties in the \teff\, values were estimated by introducing an  1$\sigma$ change in the 
%normalization level and considering 0.2\,dex error in \logg\, assumption.
%were estimated changing the considering parameters 
%and controling the 1-$\sigma$ change in \teff. There is also possible errors caused by the uncertainties in the initial \vsini\, 
%values but this error contribution comes from the initial \vsini\, value is negligible.} 
These uncertainties were summed in quadrature with the errors obtained by the
fitting procedure. The final results for \teff\, values and their errors are reported in Table\,\ref{initialvalues}.

\begin{figure*}
\centering
\includegraphics[width=18cm,angle=0]{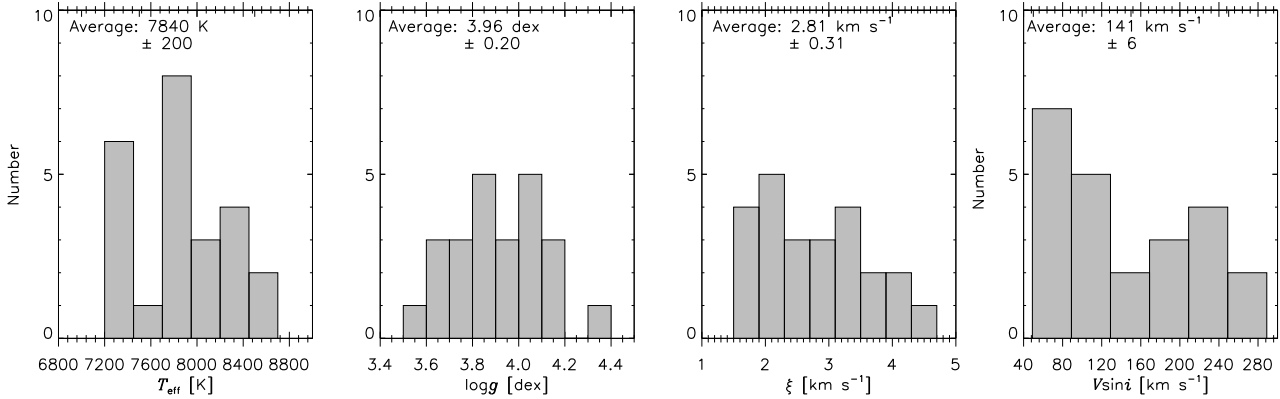}
\caption{Distributions of the final atmospheric parameters determined from the iron lines analysis and \vsini\, values.}
\label{NOT_paramater_dist}
\end{figure*}

Final atmospheric parameters were derived by using the 
excitation and the ionisation potentials of metal lines. For the correct atmospheric parameters of a star, 
all lines of the same element should give the same chemical abundance. The relationship between 
the chemical abundance and the excitation, ionisation potentials of the 
same element should be flat. 
%\textbf{The methodology used in 
%this study have been used in the several studies \citep[e.g.][]{2017MNRAS.470.2870N, 2012MNRAS.422.2960T, 2010EAS....43..167N}. 
%To obtain more reliable atmospheric parameters, we analysed individual lines or very narrow spectral parts ($\sim$0.5\,$-$\,1 nm), 
%instead of analysing entire spectrum.}

We used Fe lines in the analysis, since they are the most numerous lines %lines of this element are more available 
in the spectra having the \teff\, range of our stars. The ATLAS9 \citep{1993KurCD..13.....K, kur93} model atmospheres were synthesized by using the SYNTHE code \citep{1981SAOSR.391.....K} 
in this and the following chemical abundance analysis. %The spectrum synthesis method was applied in the research. In this method, 
 A synthetic spectrum is adjusted until it fits well the observed spectrum looking at $\chi^2$ parameter (for more details see
\citeauthor{2015MNRAS.450.2764N} \citeyear{2015MNRAS.450.2764N},
\citeauthor{2016MNRAS.458.2307K} \citeyear{2016MNRAS.458.2307K}).
The Fe lines of all stars were analysed for a range of \teff, \logg, and $\xi$ with a step of 
100\,K, 0.1\,dex and 0.1\,\kms, respectively. The range of the atmospheric parameters was 
selected taking into account the initial \teff\, values derived from Balmer lines 
and the values given by M17 (Table\,\ref{initialvalues}). After the analysis 
was performed, we determined \teff\, and \logg\, values considering the excitation potential$-$abundance and 
the ionisation potential$-$abundance relations, respectively. The $\xi$ values were also obtained by checking the dependence 
between abundance and line strength. The obtained  atmospheric parameters are given in Table\,\ref{Feresult}. The 
uncertainties in the parameters were estimated by checking how much the parameters change for $\sim$$5$\,\% differences in the excitation potential$-$abundance, 
ionisation potential$-$abundance, and the abundance$-$line strength relationships.

\begin{figure}
 \centering
 \begin{minipage}[b]{0.43\textwidth}
  \includegraphics[height=3.7cm, width=0.9\textwidth]{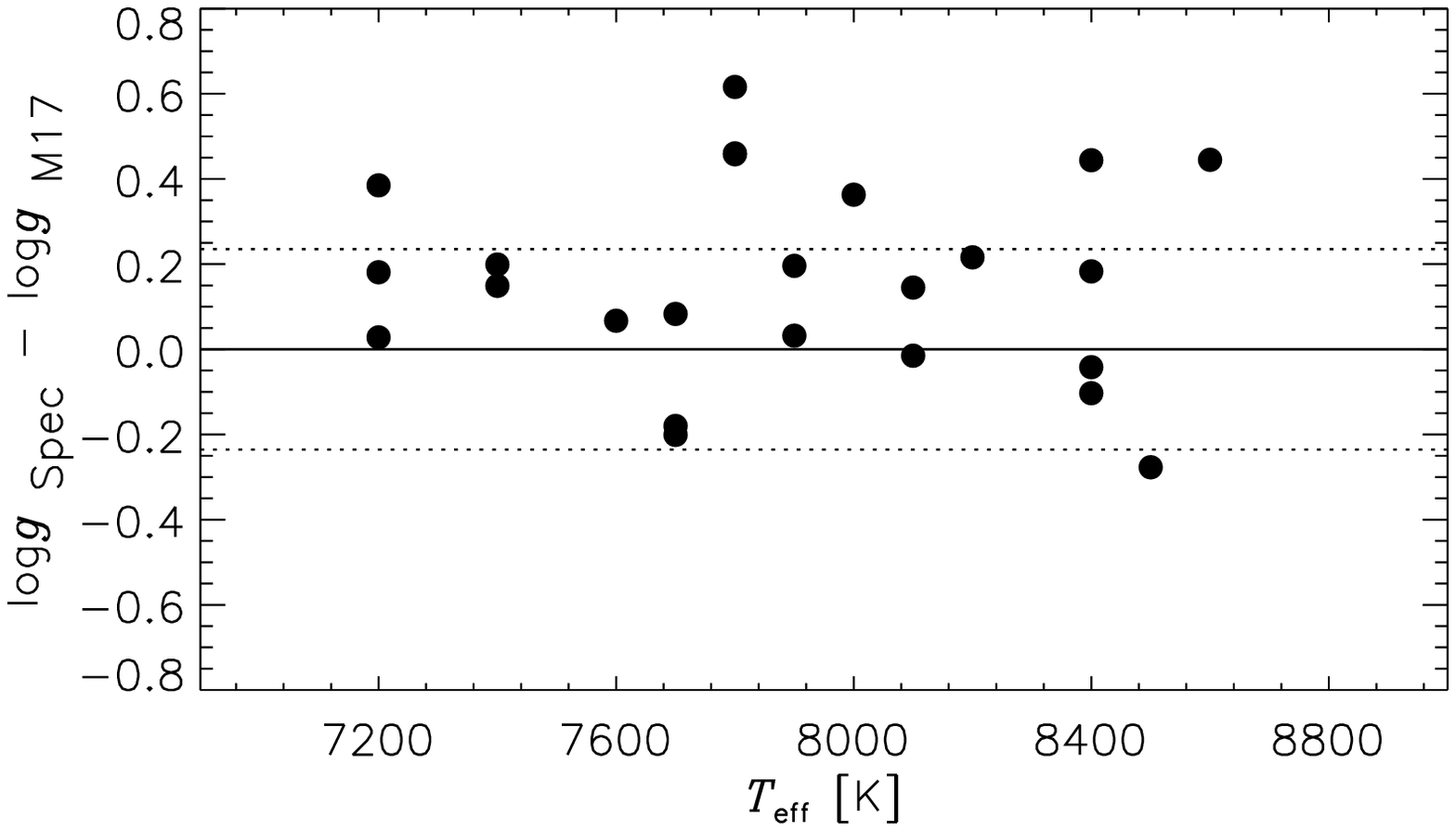}
  \end{minipage}
 \begin{minipage}[b]{0.43\textwidth}
  \includegraphics[height=3.7cm, width=0.9\textwidth]{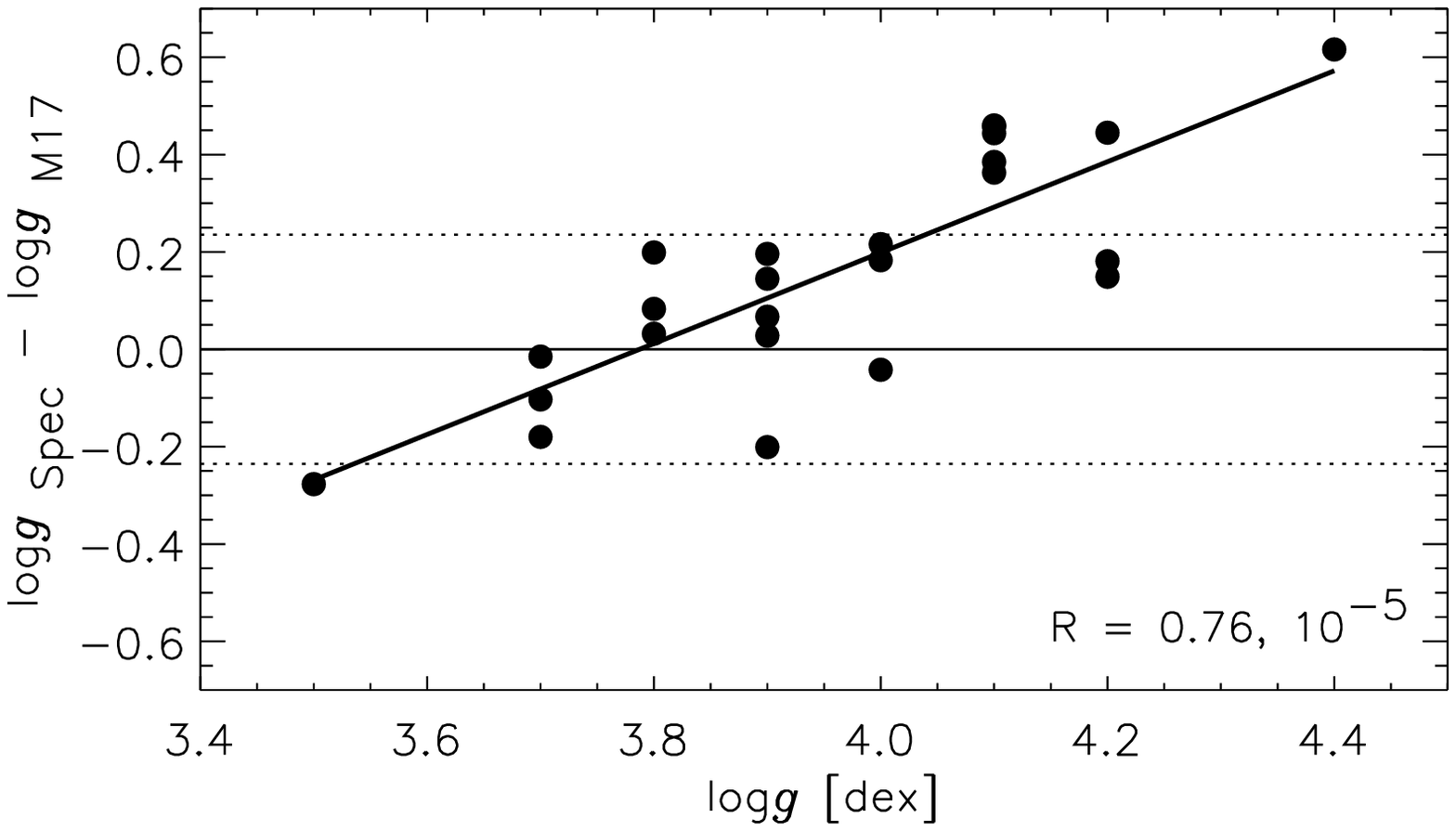}
%   \caption{3}
 \end{minipage}
   \caption{Differences between the final spectroscopic \logg\, and the \logg\, taken from M17 as a function
of the final spectroscopic \teff\, (upper panel) and \logg\, (lower panel) obtained from the iron line analysis. Dashed lines represent 1-$\sigma$ levels. 
The Spearman's rank correlation coefficient (R) and probability (the number after comma) are given in the right corner of the right panel.}
%In below panel, the first number of R constant represents the strength of the correlation (in the ideal case close to 1) 
%while the second number illustrates the deviations of points from the correlations (in the ideal case close to 0).}
 \label{NOT_FeTeff-logKIC}
\end{figure}

The distributions of the derived atmospheric parameters are shown in Fig.\,\ref{NOT_paramater_dist}. 
The final \teff, \logg, and $\xi$ ranges were obtained to be $7200-8600$\,K, $3.5-4.4$\,dex, and $1.5-4.4$\,\kms, 
respectively. The final atmospheric parameters were compared with the M17 atmospheric 
parameters. In most cases \teff\, values are consistent with each other within errors. 
However, in Fig.~\ref{NOT_FeTeff-logKIC}, the final \logg\, values were also compared 
with the \logg\, values given by M17. The final \logg\, values are 
generally higher than the M17 \logg\, values for the \teff\, ranges of our stars. Additionally, 
it appears that the \logg\, difference between 
our spectroscopic values  and M17 ones increases with the growing \logg. 
The relationship between those parameters is shown in the lower panel of Fig.~\ref{NOT_FeTeff-logKIC}.

% \citet{2017MNRAS.470.2870N} 
% showed the same figure with a wider sample. Similar to our result, they showed the spectroscopic \logg\, values are  
% mainly higher than the \logg\, given in M17 for the same \teff\, range. 

% \begin{figure}
% \includegraphics[width=8cm,angle=0]{NOT_FeTeff-logKIC.ps}
% \caption{Differences between the final spectroscopic \logg\, and the \logg\, taken from M17 as a function
% of the final spectroscopic \teff\, obtained from the iron line analysis. Dashed lines represent 1-$\sigma$ levels.}
% \label{NOT_FeTeff-logKIC}
% \end{figure}

\begin{figure}
\includegraphics[width=8.5cm,angle=0]{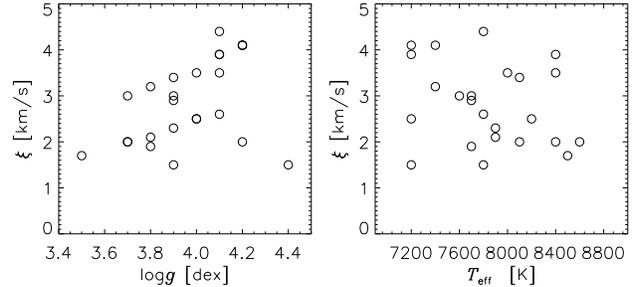}
\caption{The final $\xi$ as a function of the final spectroscopic \teff\, (right panel) and \logg\, (left panel).}
\label{NOT_mic-_Teff_log}
\end{figure}

Relations between $\xi$ and \teff\, and \logg\, were checked as shown in Fig.\,\ref{NOT_mic-_Teff_log}. 
The $\xi$\,$-$\,\teff\, relationship has already been examined in several studies (e.g.,
 \citealp{2009A&A...503..973L, 2014psce.conf..193G, 2017MNRAS.470.2870N}). 
% \citealp{2017MNRAS.470.2870N, 2014psce.conf..193G, 2009A&A...503..973L}). 
According to these studies, a decline in $\xi$ is expected for the \teff\, value higher than about 7000\,K. 
% The obtained $\xi$\,$-$\,\teff\, relation in this study is consisted with the previously given ones. 
The $\xi$\,$-$\,\logg\, relationship was also examined by \citet{2001AJ....121.2159G}. They showed a relation between these 
two parameters for spectral types from  A5 to G2. According to this relation, the  $\xi$ values decrease with increasing \logg. 
However, as our sample spans a  \teff\, range narrower than the A5\,$-$\,G2 one, 
we could not verify these relationships (Fig.\,\ref{NOT_mic-_Teff_log}).

\begin{table*}
\centering
\footnotesize
  \caption{The final atmospheric parameters derived from the iron lines analysis, the \vsini\, values and the Fe abundances. The final pulsation type classification 
  is given in last column as a result of our frequency analysis. }
  \label{Feresult}
  \begin{tabular*}{0.95\linewidth}{@{\extracolsep{\fill}}lrccccrr@{}}
\toprule 
 Number & KIC   &  \teff            & \logg           & $\xi$            &$\log \epsilon$ (Fe)  & \multicolumn{1}{c}{\vsini}         & \multicolumn{1}{r}{Pulsation}     \\
        &       &      [K]          &  [dex]          &    [\kms]        &[dex]                 &  [\kms]        & \multicolumn{1}{r}{type}\\   
%        &       &                   & 		      &                  &                      &                &\\
\midrule
    1    &2168333 & 8400\,$\pm$\,200  & 3.7\,$\pm$\,0.2 & 2.0\,$\pm$\,0.5  & 7.47\,$\pm$\,0.27    & 175\,$\pm$\,9  & \dst\,\\
    2    &3119604 & 7700\,$\pm$\,200  & 3.9\,$\pm$\,0.3 & 2.9\,$\pm$\,0.2  & 7.23\,$\pm$\,0.34    & 92\,$\pm$\,5   & \dst\,\\
    3    &3231406 & 7900\,$\pm$\,100  & 3.8\,$\pm$\,0.2 & 2.1\,$\pm$\,0.2  & 7.63\,$\pm$\,0.31    & 169\,$\pm$\,10 & Hybrid\\
    4    &3240556 & 7800\,$\pm$\,200  & 4.4\,$\pm$\,0.2 & 1.5\,$\pm$\,0.5  & 7.51\,$\pm$\,0.35    & 213\,$\pm$\,9  & \dst\,\\
    5    &3245420 & 8100\,$\pm$\,200  & 3.7\,$\pm$\,0.1 & 2.0\,$\pm$\,0.2  & 7.58\,$\pm$\,0.32    & 151\,$\pm$\,6  & \dst\,\\
    6    &3868032 & 8400\,$\pm$\,200  & 4.0\,$\pm$\,0.3 & 3.5\,$\pm$\,0.4  & 7.11\,$\pm$\,0.32    & 181\,$\pm$\,9  & Non-pulsator\\
    7    &4677684 & 8500\,$\pm$\,200  & 3.5\,$\pm$\,0.2 & 1.7\,$\pm$\,0.3  & 7.52\,$\pm$\,0.28    & 71\,$\pm$\,3   & \gd\,\\
    8    &4768677 & 8600\,$\pm$\,300  & 4.2\,$\pm$\,0.3 & 2.0\,$\pm$\,0.5  & 6.94\,$\pm$\,0.50    & 256\,$\pm$\,14 & \dst\,\\
    9    &5180796 & 8400\,$\pm$\,200  & 4.0\,$\pm$\,0.2 & 2.5\,$\pm$\,0.2  & 7.73\,$\pm$\,0.30    & 152\,$\pm$\,6  & \gd\,\\
   10    &5630362 & 7700\,$\pm$\,300  & 3.7\,$\pm$\,0.3 & 3.0\,$\pm$\,0.5  & 7.34\,$\pm$\,0.33    & 230\,$\pm$\,12 & \gd\,\\
   11    &6199731 & 7800\,$\pm$\,200  & 4.1\,$\pm$\,0.3 & 4.4\,$\pm$\,0.4  & 7.23\,$\pm$\,0.35    & 236\,$\pm$\,10 & \dst\,\\
   12    &6500578 & 7700\,$\pm$\,300  & 3.8\,$\pm$\,0.2 & 1.9\,$\pm$\,0.3  & 7.35\,$\pm$\,0.30    & 95\,$\pm$\,6   & \gd\, \\
   13    &6776331 & 7900\,$\pm$\,200  & 3.9\,$\pm$\,0.1 & 2.3\,$\pm$\,0.2  & 7.65\,$\pm$\,0.32    & 49\,$\pm$\,3   & Hybrid\\
   14    &7694191 & 8400\,$\pm$\,200  & 4.1\,$\pm$\,0.2 & 3.9\,$\pm$\,0.2  & 7.31\,$\pm$\,0.36    & 76\,$\pm$\,6   & \gd\,\\
   15    &7732458 & 7800\,$\pm$\,200  & 4.1\,$\pm$\,0.2 & 2.6\,$\pm$\,0.2  & 7.73\,$\pm$\,0.34    & 87\,$\pm$\,5   & \dst\,\\
   16    &9052363 & 8000\,$\pm$\,200  & 4.1\,$\pm$\,0.2 & 3.5\,$\pm$\,0.3  & 7.40\,$\pm$\,0.33    & 109\,$\pm$\,5  & Non-pulsator\\
   17    &9775385 & 7400\,$\pm$\,200  & 4.2\,$\pm$\,0.2 & 4.1\,$\pm$\,0.2  & 7.44\,$\pm$\,0.36    & 71\,$\pm$\,3   & Hybrid\\
   18    &10281360& 7200\,$\pm$\,200  & 4.2\,$\pm$\,0.2 & 4.1\,$\pm$\,0.3  & 7.29\,$\pm$\,0.36    & 108\,$\pm$\,5  & \gd\,\\
   19    &11197934& 7600\,$\pm$\,200  & 3.9\,$\pm$\,0.2 & 3.0\,$\pm$\,0.5  & 7.46\,$\pm$\,0.31    & 267\,$\pm$\,18 & Hybrid\\
   20    &11199412& 7200\,$\pm$\,200  & 4.1\,$\pm$\,0.2 & 3.9\,$\pm$\,0.3  & 6.68\,$\pm$\,0.35    & 77\,$\pm$\,6   & \gd\,\\
   21    &11508397& 7200\,$\pm$\,200  & 3.9\,$\pm$\,0.3 & 1.5\,$\pm$\,0.5  & 7.68\,$\pm$\,0.30    & 240\,$\pm$\,11 & Hybrid\\
   22    &11612274& 7400\,$\pm$\,200  & 3.8\,$\pm$\,0.1 & 3.2\,$\pm$\,0.2  & 7.38\,$\pm$\,0.27    & 130\,$\pm$\,7  & \gd\,\\
   23    &11718839& 8100\,$\pm$\,200  & 3.9\,$\pm$\,0.1 & 3.4\,$\pm$\,0.2  & 7.45\,$\pm$\,0.32    & 57\,$\pm$\,3   & \dst\,\\
   24    &11822666& 8200\,$\pm$\,200  & 4.0\,$\pm$\,0.1 & 2.5\,$\pm$\,0.2  & 7.37\,$\pm$\,0.31    &115\,$\pm$\,7   & Hybrid\\
\bottomrule
\end{tabular*}
\end{table*}

% \begin{figure}
% \includegraphics[width=8cm,angle=0]{NOT_vsni.ps}
% \caption{Distribution of the \vsini\, values.}
% \label{NOT_vsni}
% \end{figure}

\section{Analysis of chemical abundances}\label{anche}

Chemical abundances of the stars were derived by performing spectrum synthesis. The derived atmospheric parameters 
were taken as input during the analysis and the chemical abundances of individual elements were determined in addition to \vsini. 

In the first step of the analysis, all lines in the spectra were divided into subsets line by line. In the case of rapidly 
rotating stars, lines are mostly blended. Therefore, for rapidly rotating stars wider ranges in wavelength were selected 
considering the normalisation level. For each subset the line identifications were done by using the line 
list of Kurucz\footnote{kurucz.harvard.edu/linelists.html}. Then these spectral subsets were analysed separately. 
The identified elements in each spectral subset and \vsini\, values were 
adjusted during the analysis. Taking the minimum 
difference between the observed and theoretical spectra, chemical abundances and \vsini\, values 
were obtained from each spectral subset. 
%The final \vsini\, values are also listed in Table\,\ref{Feresult}. 
The range of \vsini\,  was found to be from 49 to 267~\kms\, 
(Table~\ref{Feresult}, Fig.\,\ref{NOT_paramater_dist}). 

The average values of the abundances of individual elements for each star 
are given in Table\,\ref{abundances}
%as shown in Fig.\,\ref{NOT_paramater_dist}. 
%There is no slowly rotating star (\vsini\,$<$\,50\kms) in our sample. 
and the uncertainties given in this table 
are the standard deviations. The total uncertainties were estimated by considering the 
errors in the obtained atmospheric parameters, assumptions in the model atmospheres, the resolution and the S/N ratio of spectra
\citep{2016MNRAS.458.2307K}.
%In this estimation the same method performed by \citet{2016MNRAS.458.2307K} was used. 
As a result, the uncertainty in the 
obtained abundances was found to be 0.28\,dex on average. The total uncertainties were estimated for the Fe abundances (Table\,3).

\begin{figure}
\includegraphics[width=8.5cm,angle=0]{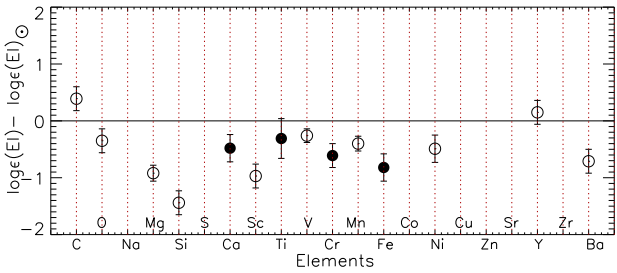}
\caption{Differences between the derived chemical abundances and the Solar values \citep{2009ARA&A..47..481A} as a function of elements for KIC\,11199412. 
Filled circles 
show the differences that were obtained from at least 5 and more lines, while others represent the opposite.}
\label{KIC11199412_NOT_abundance_dis}
\end{figure}

\begin{figure}
\includegraphics[width=8.5cm,height=3.5cm, angle=0]{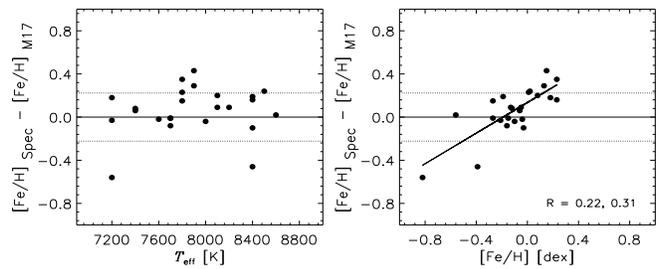}
\caption{Differences between the spectroscopic [Fe/H] and [Fe/H] taken from H14 as a function
of the spectroscopic [Fe/H] (left panel) and \teff\, (right panel). The Spearman's rank correlation coefficient (R) and probability (the number after comma) are given 
in the right corner of the right panel.}
% the first number of R constant represents the strength of the correlation (in the ideal case close to 1) 
%while the second number illustrates the deviations of points from the correlations (in the ideal case close to 0).}
\label{NOT_FeHcomparison_2figures}
\end{figure}

%The range of \vsini\, was found from $\sim$50 to 267 \kms\, as shown in Fig.\,\ref{NOT_paramater_dist}. 
%There is no slowly rotating star (\vsini\,$<$\,50\kms) in our sample. 
%Approximately $50$\,\% of the stars have \vsini\, values higher than average \vsini\, value in this study ($\sim$141 \kms). 
%The Fe abundances of the stars are also listed in Table\,\ref{Feresult}. 
Consequently, we found that all stars have chemical 
abundances similar to solar \citep{2009ARA&A..47..481A}. There are three stars, KIC\,3868032, KIC\,4768677 and KIC\,11199412, 
which seem to have moderately underabundant Fe ([Fe/H]\,$\lesssim$\, $-$\,0.50\,dex). However, when the uncertainties in Fe abundances 
of these stars were considered, we can say that only KIC\,11199412 has slightly lower Fe abundance. Additionally, this star displays 
underabundance in almost all elements compared to the solar abundances
%. The abundance pattern of this star is shown in 
(Fig.\,\ref{KIC11199412_NOT_abundance_dis}). 
We compared our [Fe/H] values and the values given by M17 (Fig.\,\ref{NOT_FeHcomparison_2figures}). 
According to the right panel, the difference between the spectroscopic [Fe/H] and the [Fe/H] values given by M17 increases 
after [Fe/H]=0.0\,dex. However, the Spearman's rank coefficient R do not support the significance of such a trend:
the correlation coefficient of $R=0.22$ and a probability value 0.31 indicate that there is no significant correlation.

Finally, KIC\,11199412 clearly shows weak
metal and \ion{Ca}{ii}\,K lines. %There is also another star, 
KIC\,11718839 also  exhibits
slightly weaker metal and \ion{Ca}{ii}\,K lines comparing the hydrogen spectral type,
but not enough to classify it as a peculiar star.
% However, the difference in the
%spectral type of hydrogen, metal and \ion{Ca}{ii}\,K lines is not enough to classify the star as a peculiar object. 
For sake of completeness, we report our final supervised spectral classification in
 Table~\ref{initialvsini}.

\section{Frequency analysis of photometric data}\label{freana}

To precisely classify the pulsational behaviour of the selected targets, we
performed an independent  frequency analysis of the {\it Kepler} data.
The original time series  were retrieved from the {\it MAST} archive\footnote{http://archive.stsci.edu/}. Keeping the
subdivision  into long- and short-cadence acquisition mode, the original data were normalized to the mean
values of each quarter, thus correcting instrumental drifts. During this procedure isolated outliers were removed.
When reconstructing the pulsational content of our variables, we used the short-cadence time series to investigate
the region above the Nyquist frequency ($f$\,=\,24.5~\cd) of the more numerous long-cadence data.

The accurate time series  were then analysed for their constituent frequencies with the goal to establish their modal content.
We have used the iterative both sine-wave fitting method (Vani\^cek 1971) and the  software package Period04 \citep{2005CoAst.146...53L}.
%computing the reduction factor of the initial variance for each trial frequency.
The final goal is to determine the pulsational characteristics with respect to the spectroscopic properties
and the position in the H-R diagram. The initial variability classification of our targets (\gd, hybrid) was
taken from \citet {2011A&A...534A.125U}. When our analysis was almost finished,  \citet{2019MNRAS.485.2380M} used {\it Gaia}--derived
luminosities to propose another classification scheme based on the skewness of the amplitudes in the Fourier spectra.
However, these authors focused on the \dst\, stars, without investigating  the frequency region below 5~\cd, where \gd\, and
hybrid stars are expected to show their $g$-modes.

\subsection{Variability not induced by pulsation}\label{rotat}

The most obvious case was that of KIC~3868032. The frequency spectrum clearly shows peaks at $f$\,=\,0.40~\cd, 2$f$, 3$f$,
and 4$f$. The spectroscopic analysis pointed out a mean profile with two superimposed components, with clearly
different rotational regimes, i.e., \vsini\,=\,180~\kms\, and \vsini\,$<100$~\kms. 
%revealed that the star is a single-lined binary.
\citet {2011A&A...534A.125U} classified it as a \gd\, variable.
However, these peaks are associated with the orbital motion and cannot be ascribed to pulsation.
We noticed that after considering $f$ and harmonics,
a peak at $f_1$\,=\,1.68~\cd\, appears, close but not equal to 4$f$. It could be still an artifact of the
orbital/rotational effects, but its pulsational origin  cannot be ruled out.

The case of KIC~9052363 is similar. The light curve is almost flat, without any clear feature that could be ascribed
to pulsation. The frequency analysis reveals a couple of low-frequency, very-low amplitude peaks. Due to
their incoherence, small rotational effects  are the most plausible reason. 
%Its classification as a hybrid variable seems inconsistent.
%Both these non-pulsating stars are reported with a ``0" flag by \citet{2019MNRAS.485.2380M}.
Both stars are also reported as non-\dst\, by \citet{2019MNRAS.485.2380M}.

%\subsection{Monoperiodic variables}

\begin{figure}
% \vspace{-3cm},
% \includegraphics[width=18cm,angle=0]{gdorall3.ps}
%%BoundingBox: 40 144 324 718
%\includegraphics[width=8cm]{gdorall5.ps}
\includegraphics[width=8.5cm]{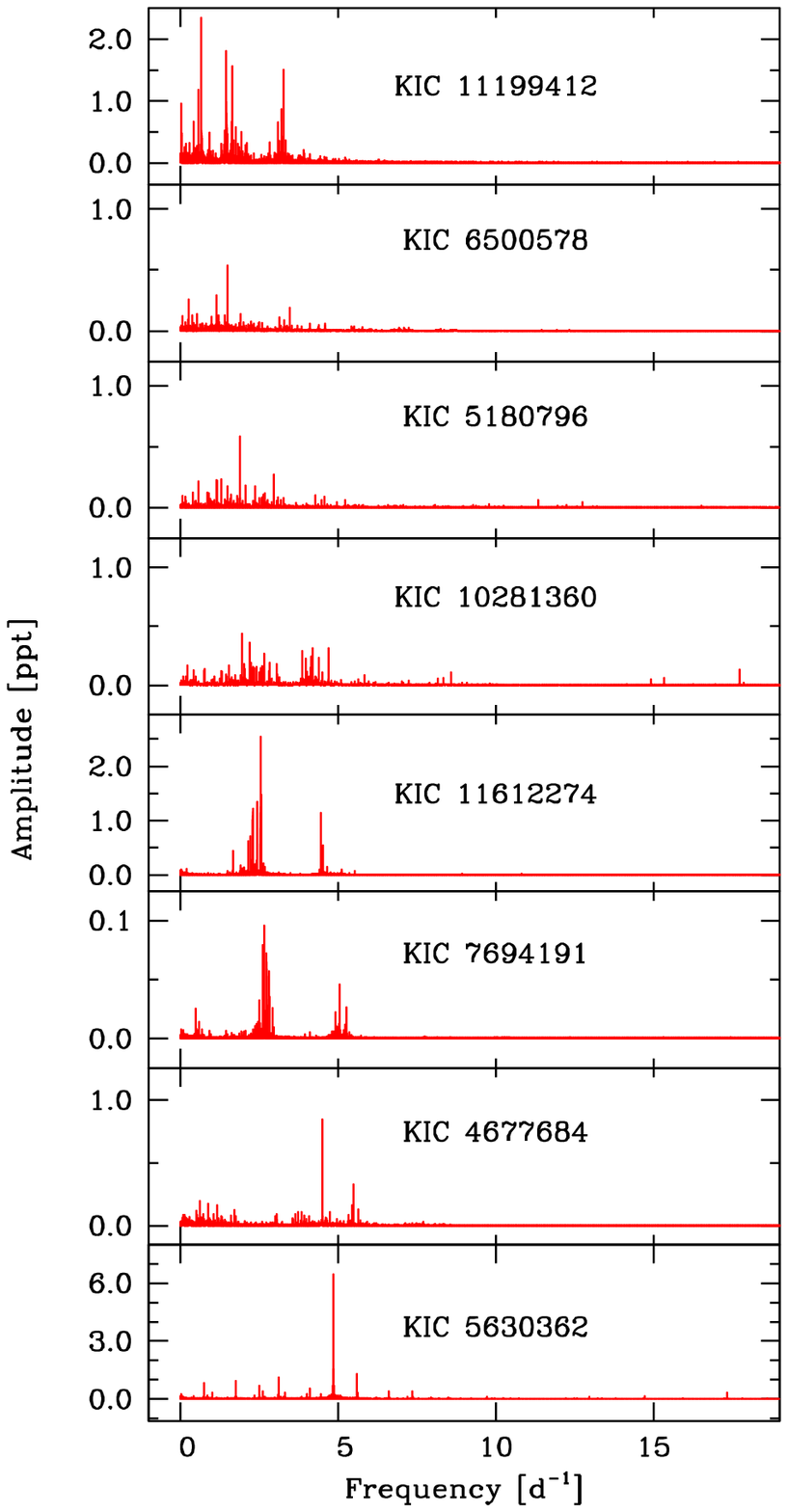}
% \vspace{-5cm}
\caption{Amplitude  spectra obtained by combining short- and long-cadence data of the variables showing 
a regime of low-frequency ($f\le6$~\cd) modes. They are ordered for increasing frequency of the highest peak.}
%Only in the case of KIC~10281360 an isolated, small-amplitude high-frequency ($f\sim18$~\cd) peak is detected.}
\label{gdor}
\end{figure}

\subsection {Low-frequency regime}\label{lowf}

Eight stars in our sample show a set of low-frequency peaks only. We ordered them by increasing frequency
of the strongest peak (Fig.~\ref{gdor}). We note that the range of frequencies is small, not exceeding
6~\cd. Occasional high frequency peaks appear in the frequency spectra not enough to claim that a clear double regime of pulsation is present. 
It is noteworthy that all stars in Fig.~\ref{gdor}\, were classified as \gd\, by \citet{2011A&A...534A.125U}.
Actually, they are the only stars classified as \gd\, in our sample and therefore we are in full agreement.
% on the pulsational content.

KIC~5630362 is the star showing the highest frequency ($f$\,=\,4.856~\cd, $P$\,=\,0.206~d), with an amplitude much
larger than those of the others (bottom panel).
% The star with the highest frequency ($f$=4.856~\cd, $P$=0.206~d), KIC~5630362, shows a has a predominant 
%mode accounting most of the variance (bottom panel). 
It is unlikely that this frequency is that of the %KIC~5630362 is pulsating in the 
fundamental radial mode, since the long period would suggest  an evolved \dst\, star: the very fast rotation
(230~\kms) and the gravity (\logg\,=\,3.7~dex) do not support such an hypothesis. Therefore, pulsators with 
largely predominant modes also exist in \gd\, stars, not only in \dst\, ones (i.e., the
Group A proposed by  \citealt{2019MNRAS.485.2380M}).

%It seems that highest is the frequency lowest is the contribution from other peaks. 

%Due to this, it should be put in the Group A proposed by \citet{2019MNRAS.485.2380M}
%belong to Group A 
%(see the KIC 10079420 case in the top panel of Fig.~B2 in \citealt{2019MNRAS.485.2380M}), but with the 
%important remark that such monoperiodic pulsators are \gd\, stars, not \dst\, ones.

\begin{figure}
% \vspace{-3cm}
% \includegraphics[width=18cm,angle=0]{dsct3.ps}
\includegraphics[width=8.5cm]{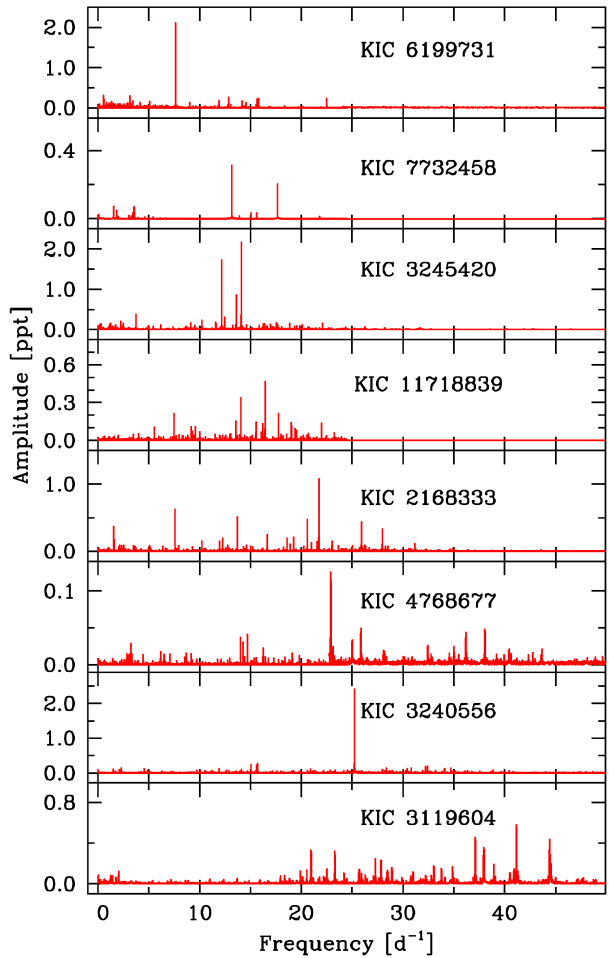}
%%BoundingBox: 30 179 323 641
% \vspace{-5.5cm}
\caption{Amplitude spectra obtained by combining short- and long-cadence data of the variables showing a 
prevailing regime of high-frequency modes. They are ordered for increasing frequency of the highest peak.
%Low-frequency peaks are detected in some cases, without characterizing an effective regime.
}
\label{dsct}
\end{figure}

\subsection {High-frequency regime}

None of the stars in our sample was classified as a pure \dst\, star. However, our frequency analysis
showed that some stars are characterized by a large set of high-frequency peaks accompanied by no
(or a few) low-frequency ones (Fig.~\ref{dsct}).
The case of KIC~3240556 is noteworthy, since this star shows a largely predominant mode.
%seems really monoperiodic.  
The frequency is too high ($f$\,=\,25.2~\cd) to be due to rotation. Note that  the short-cadence
time series allowed us to determine the exact value of this high frequency very close to the
Nyquist frequency of the long-cadence data.
% without the ambiguities introduced by the Nyquist frequency in the long-cadence data. 
KIC~3240556 surely belongs to Group A in the \citet{2019MNRAS.485.2380M} classification scheme.

All these variables were classified as hybrid stars. However, 
%sparse and small-amplitude low-frequency modes can be due to rotational spliting.
we know from the analysis
of line-profile variations that very high-degree modes (up to $\ell$=14) are excited in \dst\, stars and they
are spectroscopically detectable \citep{2009A&A...506...85P,2012A&A...542A..24M}. 
The rotational splitting can shift the frequencies of retrograde modes of multiplets
toward low values, thus producing the bunch of peaks observed there in hybrid stars.
They also could be combination terms between high-frequency modes 
%or retrograde-modes seen as low-frequencies ones in the observer's reference frame 
or again effects of the rotational modulation induced by changing spots and/or faculae on the stellar 
surfaces.  We emphasize that in general
this multiplicity of causes does not make a few low-frequency
peaks on their own a sufficient criterion to claim for an hybrid pulsational regime.
%We emphasize that  in %The decrease from 100\% to 33\% 
%out of the initial sample of 24 (33\%), can help us to understand the pulsations in those hot stars and the origin of them, even the number of sample is small. 
%%\ep {They constitute our bona-fide sample of hot pulsators, both hybrid and \gd\, ones.}

%This smaller number of stars is still important, but
%the lesser impact suggests
%that particular physical conditions, as those recalled in Sect.~1, can make a significant contribution to explaining the
%puzzle. Let us quicky go through them.
%In the next sections, we will discuss the existence of hot \gd\, and hybrid stars by taking into account the binary nature, rotational velocity properties, and SPB hypothesis.
%
% Let us quicky go through them.
%severe problems, but suggest that actuallt invertigation  This high percentage strongly reduces the expected number of 
%Most candidate hybrid stars werV486 Scoe classified to be \dst\, variables in our sample ($\sim$67\%). Only five stars were found to be hybrid variables 
%and three of them are hot hybrid stars (\teff\,$\geq$\,7400\,K). We also determined two non-pulsators which only have a variability induced by rotation and 
%binarity. 

%Consequently, we defined the pulsation type of the targets and confirmed that there are hot \gd\, and hybrid stars. The list of the new pulsation types of the 
%stars is given in Table\,3general this multiplicity
%even if a few $g$-modes  can be excited in  high-frequency pulsators.}
%low-frequency, low-amplitude  modes can be excited in a high-frequency pulsator, 

\subsection{Low and high-frequency regime}

\begin{figure}
%  \vspace{-3cm}
% \includegraphics[width=18cm,angle=0]{hybrid3.ps}
%\includegraphics[width=8.5cm]{hybrid4.eps}
%%BoundingBox: 32 287 325 693
\includegraphics[width=8cm]{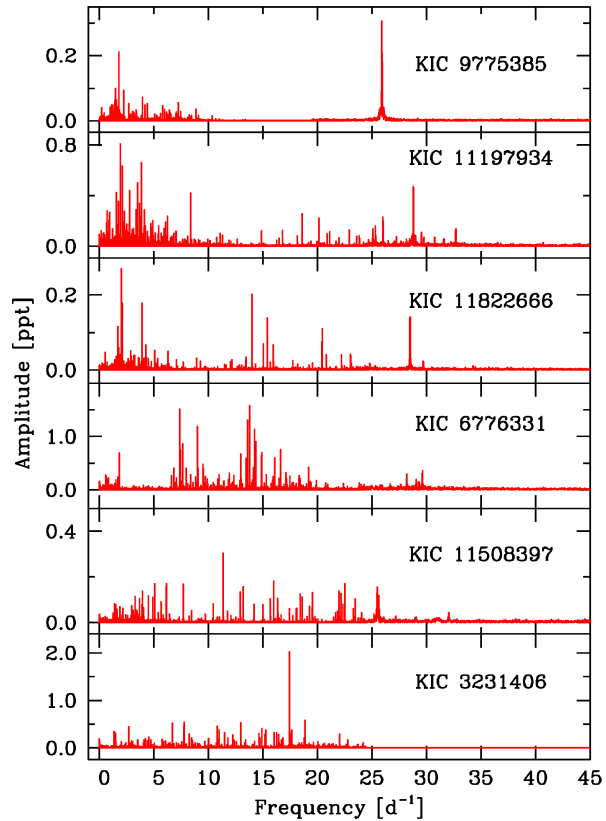}
%\vspace{-6cm}
\caption{Amplitude spectra obtained by combining short- and long-cadence data of the variables showing clear regimes of both high-frequency
and low-frequency peaks.}
\label{hyb}
\end{figure}

Only six stars show simultaneously well-defined low- and high-frequency regimes
(Fig.~\ref{hyb}), with clear peaks in both regions.
However, it has to be noted  how the amplitude spectra differ: there are 
low- and high-frequencies both confined in separated groups and continuously distributed.
%both groups of low- and high-frequencies confined in separated intervals or continuously distributed.}
%a number of low- and high-frequency peaks well distributed over a large interval
%(KIC 11508397, bottom panel), passing from intermediate peak structures (KIC 11197934,
%KIC 11822666,  central panels).
All these stars %shown in Fig.~\ref{hyb} i
had an input classification as hybrid variables.% by \citet {2011A&A...534A.125U}.

\section{Discussion}\label{disc}

\begin{figure*}
\includegraphics[width=18cm,angle=0]{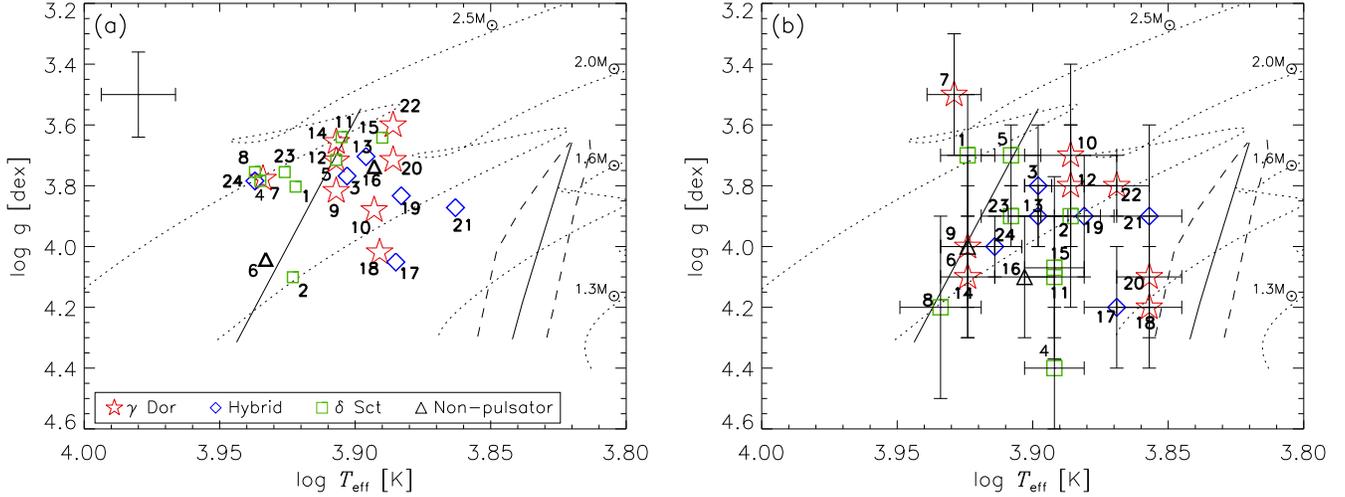}
\caption{ {\it a)} The positions of the target stars, as numbered in Table~\ref{Feresult}, 
according to the atmospheric parameters given by M17.
 {\it b)} The positions of the stars according to the final atmospheric parameters. 
%The stars are numbered considering the given number to the stars in tables. 
The theoretical instability strips of the 
$\gamma$ Dor (dashed-lines) and $\delta$ Sct (solid lines) stars were taken from \citet{2005A&A...435..927D}. 
The evolutionary tracks (Z=0.02) were 
adopted from \citet{2016MNRAS.458.2307K}.}
\label{NOT_hr_filiz_logg}
\end{figure*}

We can now use the results from the frequency and atmospheric analyses to better define
the properties of our targets and closely look at how much they are really
peculiar, as suggested by previous works. Our new classification and parameters 
are reported in Table~\ref{Feresult}.
% and used to draw a more reliable H-R diagram (Fig.\,\ref{NOT_hr_filiz_logg}).
%In this section, the listed questions in the introduction were addressed to find the answers. 
In the left panel of Fig.\,\ref{NOT_hr_filiz_logg}, the log\,\teff\,$-$\logg\, positions are shown  on the basis of
the predicted atmospheric parameters by M17, while in the right panel 
%the position of them is illustrated taken into account 
our final atmospheric parameters are used. As can be noticed, the larger changes in the positions are
mostly due to  the differences in the \logg\, values. 

%The decrease from 100\% to 33\% 
%out of the initial sample of 24 (33\%), can help us to understand the pulsations in those hot stars and the origin of them, even the number of sample is small. 
%%\ep {They constitute our bona-fide sample of hot pulsators, both hybrid and \gd\, ones.}

%This smaller number of stars is still important, but
%the lesser impact suggests
%that particular physical conditions, as those recalled in Sect.~1, can make a significant contribution to explaining the
%puzzle. Let us quicky go through them.
%In the next sections, we will discuss the existence of hot \gd\, and hybrid stars by taking into account the binary nature, rotational velocity properties, and SPB hypothesis.
%
% Let us quicky go through them.
%severe problems, but suggest that actuallt invertigation  This high percentage strongly reduces the expected number of 
%Most candidate hybrid stars werV486 Scoe classified to be \dst\, variables in our sample ($\sim$67\%). Only five stars were found to be hybrid variables 
%and three of them are hot hybrid stars (\teff\,$\geq$\,7400\,K). We also determined two non-pulsators which only have a variability induced by rotation and 
%binarity. 

%Consequently, we defined the pulsation type of the targets and confirmed that there are hot \gd\, and hybrid stars. The list of the new pulsation types of the 
%stars is given in Table\,3
\subsection{Binary nature} \label{binary}
%The binary nature of the stars was investigated. %Six \gd\, and eleven hybrid star candidates 
%We already reported on the binarity of  KIC\,3868032, an ellipsoidal variable (Sect.~\ref{rotat}).
Composite spectra suggesting double-lined spectroscopic binaries are not observed,
except for the ellipsoidal variable KIC\,3868032 (Sect.~\ref{rotat}).
Seventeen stars in our sample have at least two spectra taken on different nights.
The radial velocities  were examined: if there was a companion, the radial velocities should vary due to the orbital motion.
We did not find any large variation in the radial velocities. Nevertheless,
the very small number of observations and their very limited time coverage do not allow us to clearly detect long-period
or small-amplitude or low-inclination binary systems. 
 For example, %inside our sample, there are  two stars 
KIC\,2168333 and KIC\,11508397  are known
binaries with very long orbital periods \citep[$\gtrsim$~350~d; ][]{2018MNRAS.474.4322M}, 
but the radial velocities obtained from their two spectra differ by 2.6 and 1.5~\kms\,only, respectively.
%The radial velocities obtained from their two spectra differ by 2.6 and 1.5~\kms, respectively.
%\spb{
%We did not find any large variation in the radial velocities. Nevertheless,
%the very small number of observations and their very limited time coverage do not allow us to detect long-period
%or small-amplitude binary systems. }

%However, we did not find any large variations in the radial velocities.
%Nevertheless, it should be kept in mind that our limited observation period does not allow us to detect
%long-period binary systems. For example, inside our samples, there are two stars (KIC\,2168333 and KIC\,11508397) which are known
%binaries with very long orbital periods \citep[$\gtrsim$~350~d; ][]{2018MNRAS.474.4322M}. 
%Their radial velocities obtained from their two spectra differ by 2.6 and 1.5~\kms, respectively.

\subsection{Pulsation characteristics vs new atmospheric parameters}\label{classif}

Our detailed frequency analysis was able to clean the physical scenario since 
10 targets (8 \dst\, and 2 non-pulsating stars; Sects.~\ref{rotat} and ~\ref{lowf}, respectively) 
out of 24 (42\%) can be retired from the initial sample of \gd\, and hybrid pulsators.
In particular the 8  \dst\, stars erroneously classified as hybrid variables  %\citep{2011A&A...534A.125U} 
turned out to be normal
$p$-modes  pulsators, well inside the \dst\, instability strip. 
Additionally, the two non-pulsators remain located close to the hot border of the \dst\, domain, where 
the fraction of \dst\, pulsating stars is estimated to be 
$\sim$40\% \citep{2019MNRAS.485.2380M}.

Among the six hybrid variables, only  KIC\,11822666 is really a hot star (8200\,K),
%and 8200\,K, respectively), 
while the \teff\,  of KIC\,9775385, KIC\,11197934, and KIC\,11508397 are very close or below the 7300\,K limit when
taking into account the error bars. KIC~6776331 and KIC~3231406 have intermediate values (7900~K).

A similar count applies to the 8 targets re-classified as pure \gd\, stars: five 
(KIC\,5630362, KIC\,6500578, KIC\,10281360, KIC\,11199412, and KIC\,11612274) have
 \teff\, values in agreement with that of the hot border of the classical \gd\, strip (assuming 1$\sigma$ error bars),
 though  KIC\,5630362 and  KIC\,6500578 seem to be more luminous than usual for \gd\, stars (Fig.\,\ref{NOT_hr_filiz_logg_sm}).
Note that the normal \gd\, pulsator KIC\,11199412 is the only star showing hints of chemical peculiarities (Sect.~\ref{chemabund}).
 In three cases (KIC\,4677684, KIC\,7694191, and KIC\,5180796) the \teff\, values are well above 8000\,K.

In the end, we have 8 stars (3 hybrid and 5 normal \gd\, pulsators)  
that show $g$-modes even if they are beyond the hot border of the \gd\, instability strip: namely, 
KIC~3231406, KIC\,6776331, KIC\,11822666, KIC\,4677684, KIC\,5180796, KIC\,5630362, KIC\,6500578, and KIC\,7694191.
They constitute our final bona-fide sample. We recall that the whole initial sample (24 stars)
was expected to be composed of such unusual \gd\, pulsators.

\subsection{Other atmospheric parameters}

When the obtained atmospheric parameters are taken into account, most stars are found that they have \teff\, values in agreement with those
given by M17. Only two stars, KIC\,10281360 and KIC\,11199412,  show spectroscopic \teff\, values about 500\,K
cooler than the previously given values by M17.  This explains why the previous classification put them so close to the hot border.

Our final \logg\, values were compared with the \logg\, values given by M17 (Fig.\,\ref {NOT_FeTeff-logKIC}).
In some cases, the final \logg\ values differ more than 0.4\,dex. Therefore, the positions of these stars change
in the log\,\teff\,$-$\logg\, diagram (Fig.\,\ref {NOT_hr_filiz_logg}). When taking into account the final positions
 in the log\,\teff\,$-$\logg\, diagram
and the evolutionary tracks, we can say that $\sim$$70$\,\% of stars have masses from 2.0 to 2.5 $M_{\sun}$, while the mass range of
the others is between 1.6 and 2.0 $M_{\sun}$.

The microturbulent velocity $\xi$ changes with \teff\, 
\citep{2001AJ....121.2159G, 2004IAUS..224..131S, 2009A&A...503..973L} 
% in the section of convection and turbulence. 
and the $\xi$ value varies
from $\sim$1.5 to 3 \kms\, for \teff\, range of about 7000\,$-$\,7300. However, three cool stars
%shown in Fig.\,\ref{NOT_hr_filiz_logg} 
(KIC\,9775385, KIC\,10281360, KIC\,11199412) have $\xi$ values between 3.9 and 4.1~\kms with a maximum uncertainty of 0.3 \kms. These $\xi$ values are high for this \teff\, range, but the three stars
seem to be normally located on the cool border of the \gd\, instability strip. For the other hotter stars
the $\xi$ range is between 1.5 and 4.5 ($\pm$\,0.3)~\kms,  as expected for A-type stars.

\begin{figure*}
\includegraphics[width=17cm]{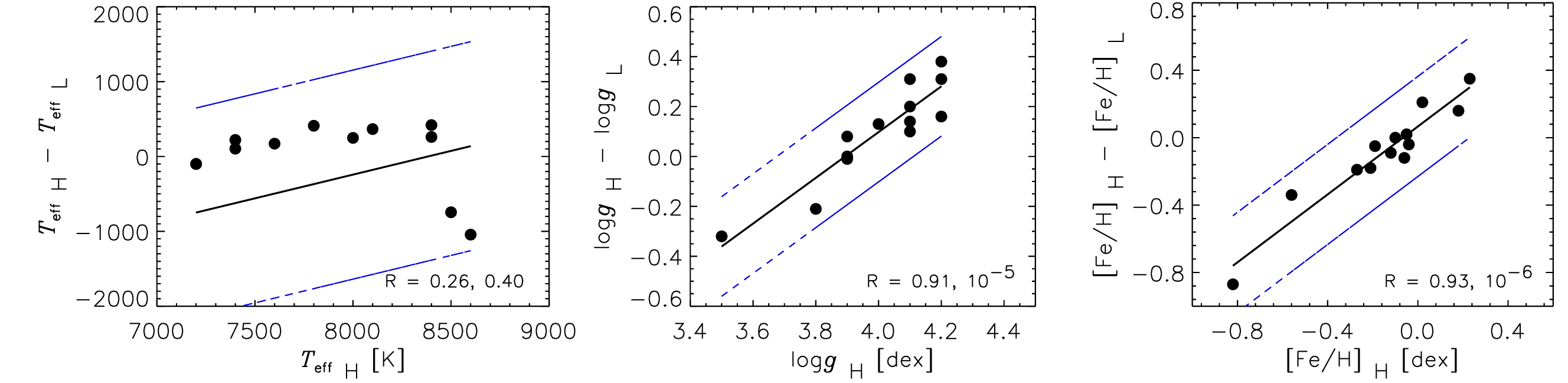}
\caption{Comparison of the atmospheric parameters obtained from high- and low-resolution spectroscopy. 
The subscripts $H$ and $L$ defines the high- and low-resolution spectroscopy, respectively. The Spearman's rank correlation coefficient (R) and probability (the number after comma) are given 
in the right corner of the panels. Blue lines represent the 1$\sigma$ levels.}
\label{lamostcomp}
\end{figure*}

Some target stars have low-resolution LAMOST spectroscopy in the literature. The atmospheric parameters derived from these 
studies are given in Table\,\ref{litatm}. When we compared these results with our final atmospheric parameters, we found significant differences 
and trends between the parameters obtained from the high- and low-resolution spectroscopy (Fig.\,\ref{lamostcomp}).

\subsection{Chemical abundances}\label{chemabund}
Anomalous abundances, like the Am phenomenon \citep{2011ApJ...743..153H},
have  been considered a possible physical
explanation of the  hybrid \gd-\dst\, pulsation.
Our analysis did not reveal any abundance peculiarity in 
our bona-fide sample of eight hot \gd\, and hybrid stars,
in agreement with recent studies showing that most hybrid stars are
chemically normal  \citep{2015MNRAS.450.2764N, 2017MNRAS.470.2870N}.
We also searched for He lines, without any significant detection.
Only KIC\,11199412 (a normal \gd\, star) exhibits a moderate underabundance 
in almost all the  elements (Fig.\,\ref{KIC11199412_NOT_abundance_dis}).
%The atmospheric chemical abundances  were obtained since  
%the Am phenomenon has been considered a possible physical
%explanation of the  hybrid \gd-\dst pulsation has
%been initially proposed \citep{2011ApJ...743..153H}% investigated 
%. 

%In the previous studies presented by 
%\citet{2015MNRAS.450.2764N, 2017MNRAS.470.2870N} also showed that most hybrid stars are chemically
%normal.

\subsection{The \vsini\, values and the SPB hypothesis} \label{spb}
It has been suggested that 
%Another explanation is that the 
hot \gd\, stars are actually rapidly-rotating SPB stars
\citep{2014A&A...569A..18S,2015MNRAS.451.1445}. 
Due to gravity darkening, their equatorial zones appear cooler than the rest of the surface, 
then they are classified as hot A-stars and, hence,
as hot \gd\, variables.

\citet{2016MNRAS.460.1318B} determined 
an average \vsini\,=\,114\,\kms\, for a group of six hot \gd\, stars. 
Our five bona-fide hot \gd\, stars  show \vsini\, values ranging from
71 to 230~\kms\, (Table~\ref{Feresult}), with an average value of 125\,\kms.
In the spectral range B5-B9, main-sequence stars, like SPB variables, show a mean \vsini\,=\,144\,\kms\,
\citep[][]{2005ESASP.560..571G, 2016MNRAS.460.1318B}. % and \teff\, values well above 10000~K. 
%For the range of spectral type of B5\,$-$\,B9 main-sequence stars, like SPB stars, show a mean \vsini\,=\,144\,\kms 
%\citep[][]{2005ESASP.560..571G, 2016MNRAS.460.1318B}.
Since  hot \gd\, stars do not seem to rotate faster than normal SPB stars, 
 they do not constitute  a special SPB subclass. 
Moreover, we could not find spectral lines indicative of the B-type (Sect.~\ref{chemabund}), which should be present at least for SPB stars seen at intermediate or pole-on orientations. Finally, gravity darkening does not
seem to be able to lower the \teff\, of SPB stars sufficiently to approach the hot border of the $\delta$ Sct instability strip
\citep[see Fig.~4 in ][]{2014A&A...569A..18S}.
We prudently note that  the classification as hot or classical 
hybrid/\gd\, stars could  also be affected by the gravity darkening, especially in case of
fast rotators seen almost equator-on, like  KIC\,5630362, KIC\,11197934, and KIC\,11508397.

\subsection { {\it Gaia} parallaxes} \label{gaia}
As a final check, we used the {\it Gaia} parallaxes \citep{2018A&A...616A...1G} to investigate the positions of the stars
in the H-R diagram. We adopted  the bolometric corrections computed taking into account the \teff\, values
\citep{1996ApJ...469..355F}. The extinction coefficients ($A$$_{V}$) were estimated using the interstellar reddening
$E(B-V)$ values (Table\,\ref{initialvalues}).
%, as obtained from the interstellar Na\,D lines. 
These were obtained by measuring the  equivalent widths of the Na\,D lines
in our spectra and then applying the relation given by \cite{1997A&A...318..269M}.

%We estimated $E(B-V)$ values from the interstellar Na\,D lines present in our spectra. The equivalent widths of the Na\,D lines were measured
%and $E(B-V)$ obtained using the relation given by \cite{1997A&A...318..269M}. The obtained $E(B-V)$ values are listed in Table\,\ref{initialvalues}.}

% the equation given by \citet{1999PASP..111...63F}.

\begin{figure}
\includegraphics[width=8cm,angle=0]{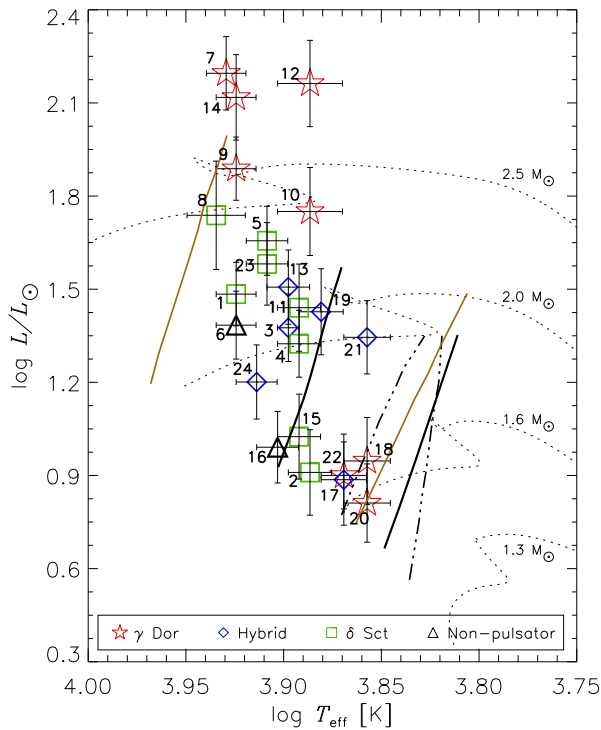}
\caption{The positions of the stars in the H-R diagram, as numbered in Table~\ref{infotable}.
% The stars are numbered considering the given number to the stars in tables. 
The theoretical instability strips of the
$\gamma$ Dor (dashed-lines) and $\delta$ Sct (solid black lines) stars were taken from \citet{2005A&A...435..927D}. The recently suggested
\dst\, instability strip \citep{2019MNRAS.485.2380M} is shown by brown solid lines. The evolutionary tracks (Z=0.02) were
adopted from \citet{2016MNRAS.458.2307K}.}
\label{NOT_hr_filiz_logg_sm}
\end{figure}

An offset of $-0.03$\,mas in the {\it Gaia} parallaxes was found \citep{2018A&A...616A...2L, 2019ApJ...878..136Z}. However,
when this offset was applied to a large sample ($\sim$\,15000 stars), a huge amount of stars
goes below the zero-age main-sequence with unreasonably low luminosities \citep{2019MNRAS.485.2380M}.
Therefore, to calculate our final Gaia luminosities we did not apply this offset to the parallaxes, accordingly with recent studies
\citep{2018A&A...616A..17A,2019MNRAS.485.2380M}. The positions of the stars in the H-R diagram are illustrated in Fig.~\ref{NOT_hr_filiz_logg_sm}.
%After the calculation of the Gaia luminosities, 
Note that the new empirical  instability strip of \dst\, stars \citep{2019MNRAS.485.2380M} now incorporates
some variables located beyond the hot border of the previous theoretical instability strip.
%do not solve the matter since the \gd stars are still located beyond the hot border of this newly suggested \dst\, instability strip.

As can be noticed from Fig.\,\ref{NOT_hr_filiz_logg_sm}, %the hot \gd\, stars (star 7, 9, 10, 12, 14) 
stars  7, 9, 10, 12, and 14 (see Table~\ref{Feresult} for their KIC identification)
have very high luminosities.
%higher luminosities comparing the \gd\, and most \dst\, stars. 
When the radii of those hot \gd\, stars were calculated using the
Gaia luminosities, we obtained values ranging from 4.3 to 6.0\,$R_\odot$.
We also calculated the radii of the hot \gd\, stars classified
by \citet{2016MNRAS.460.1318B}. It turned out that $\sim$70\% of their sample have larger radii (3.2\,$<$\,$R_\odot$\,$<$\,7.2).
Main-sequence A-F type  stars should have radii in
the range $1.5-2.7$\,$R_\odot$ \citep{2000asqu.book.....C}.
%, when we consider them as main sequence stars. 
%(typically luminosity class is IV-V for those stars). 
Therefore, the larger size of the hot \gd\, pulsators appears to  support the hypothesis that they are different from main-sequence A-F pulsators.
They could be evolved A-F stars or more massive stars (such as B-type stars) entering the classical
instability strip in their redward evolution.  
The impact of enhanced iron opacity on stellar pulsations can lead to a substantial
revision of the instability strips \citep{2016MNRAS.455L..67M} and then play a key role to
explain the location of our hot hybrid  and  \gd\, variables. Additionally, the higher luminosity values of these hot stars may be the result of binarity. 
These systems could be a member of
a long orbital period binary system which could not detect with our present
spectroscopic data.

\section{Conclusions}\label{conclu}

We performed a detailed spectroscopic and photometric study of a group of 24 {\it Kepler} targets,
%located close to the hot border of the instability strip. %pulsators hot \gd\, and hybrid stars. 
all claimed to be hot \gd\, or \dst-\gd\, hybrid stars, located well beyond the theoretical 
\gd\, instability strip \citep{2011A&A...534A.125U}.
% However, the majority of our sample consisted of stars whose \teff values were previously 
%misclassified, they actually fall within the normal \teff\, range for their group. 
The detailed frequency analysis of the {\it Kepler} time series and the 
determination of \teff\, and other atmospheric parameters  performed with the new FIES spectra allowed us to set
the bona-fide sample of such peculiar pulsators to five hot \gd\, stars and three hot hybrid stars. 
If on one hand we provide a strong confirmation that these peculiar pulsators exist,
on the other we reduce their recurrence to 1/3 to what originally evaluated. 
Therefore, we can assume that the physical explanation should reside in a mechanism or a cause applicable to 
a limited number of stars, not  to the vast majority of the \dst\, or \gd\, variables.

%with respect to the initial classification. % \citep{2011A&A...534A.125U}, 
%in the number of unusual pulsators located outside the theoretical instability strips of \gd\, and \dst\, pulsators.
Searching for this still elusive explanation, 
we did not find any peculiarity in the atmospheric parameters and chemical abundances. 
The lack of composite spectra  does not  support 
the possibility that the hot hybrid stars are binaries in which one or both components are normal \dst\, and/or \gd\, pulsators. 
We investigated the rotation rate and we cannot support the hypothesis that hot \gd\, pulsators 
are actually fast-rotating SPB stars. 
%\ep{We also prudently note that the gravity darkening  can affect the \teff\, determination of 
%hybrid and \gd\, stars, especially rapid rotators probably seen equator-on, like 
%KIC\,5630362, KIC\,11197934, and KIC\,11508397.  }
On the other hand, {\it Gaia} luminosities suggested us
that hot \gd\, pulsators have larger radii and higher luminosities than normal main sequence A-F stars.
New  efforts, both theoretical and observational,  have to be
made to well constrain all the features of this new scenario.

\section*{Acknowledgments}
The authors thank the anonymous referee for useful comments that have improved the presentation of our
results. 
The authors also wish to thank Simon Murphy for his useful comments on a first draft of the manuscript.
Based on observations made with the Nordic Optical Telescope, operated by the Nordic Optical 
Telescope Scientific Association at the Observatorio del Roque de los Muchachos, La Palma, Spain, under the 
proposal 54-003 (P.I. E.\,Poretti). The authors thank the whole NOT staff for the help in the observations, 
performed in visitor mode.
FKA and GH thank the Polish National Center for Science (NCN) for 
supporting the study through grant 2015/18/A/ST9/00578.
EN acknowledges support provided by the Polish National Science Center
through grant no. 2014/13/B/ST9/00902. 
This work has been partly supported by the Scientific and Technological Research Council of Turkey (TUBITAK) grant numbers 2214-A and 2211-C.
The calculations have been carried out in Wroc{\l}aw Centre for Networking and Supercomputing (http://www.wcss.pl), grant No.\,214. 
This research has made use of the SIMBAD data base, operated at CDS, 
Strasbourq, France.

%\appendix

\appendix

 \setcounter{table}{0}

  \begin{table*}
  \centering
  \caption{The initial \vsini\, values derived for the Balmer lines analysis 
(Sect.~\ref{atpar}) and our spectral classification (Sect.~\ref{anche}).}
 \label{initialvsini}
 \begin{tabular}{lrc}
 \toprule
    KIC   &  \vsini\,     & Spectral type\\
          &  (\kms)       \\   
  \midrule
  2168333 & 163\,$\pm$\,15 & kA5hA4\,V \\
  3119604 & 94\,$\pm$\,1 &  A5\,V \\
  3231406 & 173\,$\pm$\,6 & A7\,III-IV  \\
  3240556 & 218\,$\pm$\,4  & A5\,IV-V \\
  3245420 & 158\,$\pm$\,4   & A7\,V  \\
  3868032 & 178\,$\pm$\,4   & A5\,V     \\
  4677684 & 75\,$\pm$\,5    &  A3\,IV-V  \\
  4768677 & 257\,$\pm$\,14  &  A3\,IV-V \\
  5180796 & 155\,$\pm$\,3    &  A4\,IV-V \\
  5630362 & 242\,$\pm$\,12  & A5\,III-IV \\
  6199731 & 241\,$\pm$\,3   & A5\,III   \\
  6500578 & 98\,$\pm$\,3   & A3\,IV    \\
  6776331 & 51\,$\pm$\,1   & A7\,V   \\
  7694191 & 78\,$\pm$\,1   & A5\,IV-V    \\
  7732458 & 82\,$\pm$\,2   & A9\,IV-V   \\
  9052363 & 109\,$\pm$\,2   & A5\,IV-V   \\
  9775385 & 72\,$\pm$\,1   &   F0V   \\
  10281360& 111\,$\pm$\,4   &  A9\,IV-V \\
  11197934& 295\,$\pm$\,5   &  A8\,IV-V \\
  11199412& 73\,$\pm$\,2    &  kA3hF0mA3 V \\
  11508397& 245\,$\pm$\,6   &  F0\,V      \\
  11612274& 132\,$\pm$\,4   &  F0\,V     \\
  11718839& 57\,$\pm$\,4    &   kA7hA5mA7 V, kA6hA5mA6 V    \\
  11822666& 121\,$\pm$\,4   &  A5\,V  \\
 \bottomrule
  \end{tabular}
  \end{table*} 
  
   \setcounter{table}{1}

  \begin{table*}
  \centering
  \caption{The atmospheric parameters of the stars which have low-resolution spectroscopy in the literature. }
 \label{litatm}
 \begin{tabular*}{0.55\linewidth}{@{\extracolsep{\fill}}lrcrr}
 \toprule
    KIC   &  \teff\,         & \logg\, 		 & Fe/H               & Reference\\
          &  [K]             &  [dex]            & [dex]              &\\   
  \midrule
  4677684 & 9245\,$\pm$\,529 & 3.82\,$\pm$\,0.13 & $-$\,0.19\,$\pm$\,0.13 & 1 \\
  4768677 & 9655\,$\pm$\,323 & 3.82\,$\pm$\,0.12 & $-$\,0.22\,$\pm$\,0.12 &1 \\
  5180796 & 7981\,$\pm$\,370 & 3.87\,$\pm$\,0.13 & $-$\,0.12\,$\pm$\,0.12 &1  \\
          & 8090\,$\pm$\,10  & 3.80              & $-$\,0.04              &2  \\
  6199731 & 7390\,$\pm$\,255 & 3.96\,$\pm$\,0.13 & $-$\,0.08\,$\pm$\,0.13 &1    \\
  7694191 & 8140\,$\pm$\,10  & 3.79              & $-$\,0.14              &2     \\
  9052363 & 7752\,$\pm$\,320 & 3.90\,$\pm$\,0.12 & $-$\,0.10\,$\pm$\,0.13 &1     \\
  9775385 & 7296\,$\pm$\,137 & 3.89\,$\pm$\,0.13 & 0.06\,$\pm$\,0.14  &1     \\
  10281360& 7300\,$\pm$\,10  & 4.04              & $-$\,0.03              &2    \\
  11197934& 7429\,$\pm$\,205 & 3.90\,$\pm$\,0.12 & 0.00\,$\pm$\,0.13  &  1\\
  11199412& 11956\,$\pm$\,1854& 4.00\,$\pm$\,0.15& 0.05\,$\pm$\,0.16  &1     \\
  11508397& 7300\,$\pm$\,148 & 3.91\,$\pm$\,0.14 & 0.02\,$\pm$\,0.13  &1     \\
  11612274& 7179\,$\pm$\,152 & 4.01\,$\pm$\,0.13 & $-$\,0.03\,$\pm$\,0.13 &1  \\
          & 7175\,$\pm$\,16  & 4.07\,$\pm$\,0.02 & $-$\,0.21\,$\pm$\,0.01 &2 \\
  11718839& 7734\,$\pm$\,210 & 3.82\,$\pm$\,0.12 & $-$\,0.07\,$\pm$\,0.12 &1      \\
 \bottomrule
  \end{tabular*}
    \begin{description}
    \centering
  \item[Reference] 1. \citet{2016A&A...594A..39F}, 2. \citet{2019RAA....19....1Q}
  \end{description}
  \end{table*}

\setcounter{table}{2}
\begin{landscape}
\end{landscape}

\begin{table*}
\centering
\caption{Average abundances and standard deviations of individual elements. Number of the analysed spectral parts is given in the brackets.}
  \label{abundances}
\begin{tabular*}{0.9\linewidth}{@{\extracolsep{\fill}}lllllll}
\toprule
  Atomic  &Elements& KIC\,2168333         & KIC\,3119604          & KIC\,3231406          & KIC\,3240556          &KIC\,3245420		\\  
number    &        & 			  &                       &                       &                       &    \\
\midrule
 6 &C   &8.45\,$\pm$\,0.18 (1) & 8.54\,$\pm$\,0.19 (4) &8.53\,$\pm$\,0.13 (4)  &8.94\,$\pm$\,0.20 (2)  &8.51\,$\pm$\,0.32 (6) \\
%$_{7}$N   &                      &		         &                       &		         &			 \\
8&O   &                      &   		         &8.97\,$\pm$\,0.24  (1) &			 &                      \\
11&Na &                      &                       &5.74\,$\pm$\,0.24 (1)  &			 &                       \\
12&Mg &7.95\,$\pm$\,0.39 (6) & 7.82\,$\pm$\,0.22 (6) &8.05\,$\pm$\,0.15 (5)  &7.97\,$\pm$\,0.36 (3)  &7.85\,$\pm$\,0.50 (6)  \\
14&Si &6.60\,$\pm$\,0.18 (2) & 7.18\,$\pm$\,0.33 (3) &7.50\,$\pm$\,0.10 (3)  &6.49\,$\pm$\,0.20 (2)  &7.23\,$\pm$\,0.28 (3) \\
16&S  &			 &                       &                       &                       &                      \\
20&Ca &6.45\,$\pm$\,0.10 (3) &6.20\,$\pm$\,0.28 (12) &6.96\,$\pm$\,0.28 (7)  &6.30\,$\pm$\,0.26 (3)  &6.38\,$\pm$\,0.35 (7) \\
21&Sc &2.81\,$\pm$\,0.18 (2) &3.37\,$\pm$\,0.53 (3)  &3.63\,$\pm$\,0.32 (3)  &3.65\,$\pm$\,0.35 (3)  &3.93\,$\pm$\,0.37 (3)  \\
22&Ti &5.02\,$\pm$\,0.16 (8) &4.87\,$\pm$\,0.30 (23) &5.22\,$\pm$\,0.31 (11) &5.17\,$\pm$\,0.23 (9)  &5.40\,$\pm$\,0.45 (15)\\
23&V  &			 &3.80\,$\pm$\,0.48 (4)  &4.36\,$\pm$\,0.20 (2)  &5.35\,$\pm$\,0.23 (1)  &4.91\,$\pm$\,0.32 (2) \\
24&Cr &5.57\,$\pm$\,0.14 (6) &5.29\,$\pm$\,0.20 (12) &5.86\,$\pm$\,0.14 (8)  &5.96\,$\pm$\,0.18 (6)  &5.84\,$\pm$\,0.23 (9) \\
25&Mn &			 &4.91\,$\pm$\,0.22 (2)  &5.86\,$\pm$\,0.20 (2)  &5.45\,$\pm$\,0.22 (2)  &5.58\,$\pm$\,0.32 (2)  \\
26&Fe &7.47\,$\pm$\,0.12 (23)&7.23\,$\pm$\,0.16 (45) &7.63\,$\pm$\,0.18 (33) &7.51\,$\pm$\,0.24 (14) &7.58\,$\pm$\,0.18 (31)\\
%$_{27}$Co &                      &		         &                       &		         &                       \\
28&Ni &6.27\,$\pm$\,0.18 (1) &6.17\,$\pm$\,0.27 (3)  &6.942\,$\pm$\,0.20 (2) &5.93\,$\pm$\,0.20 (2)  &6.69\,$\pm$\,0.32 (2) \\
29&Cu &                      &		         &                       &		         &                      \\
30&Zn &                      &4.42\,$\pm$\,0.20 (1)  &		         &			 &                      \\
38&Sr &1.97\,$\pm$\,0.18 (1) &1.54\,$\pm$\,0.22 (1)  &2.11\,$\pm$\,0.20 (1)  &			 &                      \\
39&Y  &			 &4.64\,$\pm$\,0.22 (1)  &4.80\,$\pm$\,0.20 (1)  &4.03\,$\pm$\,0.32(1)   &4.66\,$\pm$\,0.32 (1) \\
40&Zr &			 &2.22\,$\pm$\,0.22 (1)  &                       & 			 &  			\\
56&Ba &2.74\,$\pm$\,0.18 (1) &1.97\,$\pm$\,0.22 (2)  &2.89\,$\pm$\,0.20 (2)  &3.27\,$\pm$\,0.32 (1)  &3.43\,$\pm$\,0.32 (1)  \\
\bottomrule
\end{tabular*}
\end{table*}

\setcounter{table}{2}

\begin{table*}
\centering
\caption{Continuation.}
\begin{tabular*}{0.9\linewidth}{@{\extracolsep{\fill}}lllllll}
\toprule
 Atomic    & Elements& KIC\,3868032         & KIC\,4677684          & KIC\,4768677          & KIC\,5180796          &KIC\,5630362		\\
 number    &        & 			  &                       &                       &                       &    \\
\midrule
6&C   &8.30\,$\pm$\,0.20 (1) & 8.60\,$\pm$\,0.27 (2) &7.99\,$\pm$\,0.13 (1)  &8.66\,$\pm$\,0.10 (5)  &8.74\,$\pm$\,0.31 (1) \\
%$_{7}$N   &                      &		         &                       &		         &			    \\
8&O   &                      & 8.64\,$\pm$\,0.27 (1) &8.31\,$\pm$\,0.24 (1)  &8.83\,$\pm$\,0.25 (1)  &                      \\
11&Na &                      &                       &			 &			 &                         \\
12&Mg &7.93\,$\pm$\,0.16 (4) & 7.85\,$\pm$\,0.33 (8) &			 &8.10\,$\pm$\,0.13 (5)  &7.87\,$\pm$\,0.51 (5)  \\
14&Si &6.42\,$\pm$\,0.20 (2) & 7.54\,$\pm$\,0.35 (7) &  			 &7.44\,$\pm$\,0.21 (5)  &6.45\,$\pm$\,0.31 (2) \\
16&S  &			 &                       &                       &                       &                      \\
20&Ca &6.57\,$\pm$\,0.20 (1) &6.25\,$\pm$\,0.11 (10) &6.96\,$\pm$\,0.28 (7)  &6.55\,$\pm$\,0.28 (9)  &5.85\,$\pm$\,0.46 (3) \\
21&Sc &3.01\,$\pm$\,0.20 (1) &3.52\,$\pm$\,0.17 (4)  &3.63\,$\pm$\,0.32 (3)  &3.49\,$\pm$\,0.32 (5)  &2.15\,$\pm$\,0.31 (1)  \\
22&Ti &4.77\,$\pm$\,0.17 (6) &5.17\,$\pm$\,0.19 (30) &5.22\,$\pm$\,0.31 (11) &5.36\,$\pm$\,0.32 (14) &4.99\,$\pm$\,0.13 (6) \\
23&V  &			 &			 &4.36\,$\pm$\,0.20 (2)  &4.81\,$\pm$\,0.25 (2)  & \\
24&Cr &5.53\,$\pm$\,0.26 (5) &5.72\,$\pm$\,0.24 (15) &5.11\,$\pm$\,0.14 (5)  &5.93\,$\pm$\,0.12 (10) &5.59\,$\pm$\,0.21 (6) \\
25&Mn &			 &5.73\,$\pm$\,0.44 (3)  &		         &5.50\,$\pm$\,0.26 (3)  &4.38\,$\pm$\,0.31 (1)  \\
26&Fe &7.11\,$\pm$\,0.16 (12)&7.52\,$\pm$\,0.16 (55) &6.94\,$\pm$\,0.18 (17) &7.73\,$\pm$\,0.17 (39) &7.34\,$\pm$\,0.23 (20)\\
%$_{27}$Co &                      &		         &                       &		         &                       \\
28&Ni &6.27\,$\pm$\,0.18 (1) &6.08\,$\pm$\,0.48 (3)  &			 &6.51\,$\pm$\,0.26 (4)  &			 \\
29&Cu &                      &		         &                       &		         &                      \\
30&Zn &                      &			 &		         &			 &                      \\
38&Sr &                      &2.61\,$\pm$\,0.27 (1)  &			 &1.99\,$\pm$\,0.25 (1)	 &                      \\
39&Y  &			 &2.22\,$\pm$\,0.27 (2)  &			 &3.83\,$\pm$\,0.25 (1)  &			 \\
40&Zr &			 &			 &                       &3.22\,$\pm$\,0.25 (1)	 &  			\\
56&Ba &1.98\,$\pm$\,0.20 (1) &2.48\,$\pm$\,0.20 (2)  &			 &2.44\,$\pm$\,0.25 (2)  &2.71\,$\pm$\,0.31 (1)  \\
\bottomrule
\end{tabular*}
\end{table*}

\setcounter{table}{2}

\begin{table*}
\centering
\caption{Continuation.}
\begin{tabular*}{0.9\linewidth}{@{\extracolsep{\fill}}lllllll}
\toprule
  Atomic    &Elements& KIC\,6199731         & KIC\,6500578          & KIC\,6776331          & KIC\,7694191          &KIC\,7732458		\\  
  number    &        & 			  &                       &                       &                       &    \\

\midrule
6&C   &8.72\,$\pm$\,0.23 (1) & 8.41\,$\pm$\,0.15 (8) &8.71\,$\pm$\,0.27 (7)  &8.76\,$\pm$\,0.27 (5)  &8.79\,$\pm$\,0.15 (7) \\
%$_{7}$N   &                      &		         &                       &		         &			    \\
8&O   &                      & 8.68\,$\pm$\,0.32 (2) &			 &			 &8.20\,$\pm$\,0.28 (2)  \\
11&Na &                      & 6.54\,$\pm$\,0.32 (1) &			 &			 &6.48\,$\pm$\,0.28 (1)  \\
12&Mg &7.00\,$\pm$\,0.23 (3) & 7.84\,$\pm$\,0.29 (8) &8.03\,$\pm$\,0.12 (7)  &7.68\,$\pm$\,0.18 (3)  &8.24\,$\pm$\,0.26 (5)  \\
14&Si &6.67\,$\pm$\,0.23 (2) & 6.97\,$\pm$\,0.48 (9) &7.07\,$\pm$\,0.32 (10) &6.90\,$\pm$\,0.49 (4)  &7.59\,$\pm$\,0.28 (10) \\
16&S  &			 &                       &                       &                       &                      \\
20&Ca &5.88\,$\pm$\,0.39 (3) &6.40\,$\pm$\,0.30 (11) &6.86\,$\pm$\,0.28 (21) &6.27\,$\pm$\,0.18 (8)  &6.73\,$\pm$\,0.35 (21) \\
21&Sc &2.69\,$\pm$\,0.23 (1) &3.12\,$\pm$\,0.35 (4)  &3.22\,$\pm$\,0.22 (6)  &3.12\,$\pm$\,0.21 (7)  &3.45\,$\pm$\,0.31 (7)  \\
22&Ti &4.97\,$\pm$\,0.25 (9) &5.00\,$\pm$\,0.23 (22) &5.43\,$\pm$\,0.22 (37) &5.23\,$\pm$\,0.25 (27) &5.45\,$\pm$\,0.31 (30) \\
23&V  &			 &3.90\,$\pm$\,0.35 (3)  &4.58\,$\pm$\,0.29 (4)  &			 &4.66\,$\pm$\,0.21 (3)\\
24&Cr &5.62\,$\pm$\,0.26 (7) &5.65\,$\pm$\,0.26 (18) &5.80\,$\pm$\,0.30 (29) &5.65\,$\pm$\,0.27 (13) &6.01\,$\pm$\,0.30 (30) \\
25&Mn &4.73\,$\pm$\,0.23 (2) &5.11\,$\pm$\,0.61 (6)  &5.46\,$\pm$\,0.23 (9)  &5.58\,$\pm$\,0.42 (6)  &5.83\,$\pm$\,0.21 (10)  \\
26&Fe &7.23\,$\pm$\,0.10 (15)&7.35\,$\pm$\,0.17 (47) &7.65\,$\pm$\,0.16 (102)&7.31\,$\pm$\,0.25 (59) &7.73\,$\pm$\,0.21 (89)\\
28&Ni &			 &6.37\,$\pm$\,0.43 (12) &6.25\,$\pm$\,0.32 (21) &6.18\,$\pm$\,0.25 (7)  &6.51\,$\pm$\,0.38 (32)\\
29&Cu &                      &		         &                       &		         &4.61\,$\pm$\,0.28 (2) \\
30&Zn &                      &			 &		         &			 &                      \\
38&Sr &                      &3.16\,$\pm$\,0.32 (1)  &3.96\,$\pm$\,0.31 (2)  &2.43\,$\pm$\,0.24 (1)	 &3.55\,$\pm$\,0.28 (2) \\
39&Y  &			 &2.34\,$\pm$\,0.32 (2)  &3.45\,$\pm$\,0.57 (3)	 &2.73\,$\pm$\,0.24 (2)  &2.53\,$\pm$\,0.20 (4) \\
40&Zr &			 &2.86\,$\pm$\,0.33 (5)  &3.40\,$\pm$\,0.76 (5)  &3.04\,$\pm$\,0.24 (1)  &3.04\,$\pm$\,0.39 (3)\\
56&Ba &2.38\,$\pm$\,0.23 (1) &2.55\,$\pm$\,0.21 (3)  &3.22\,$\pm$\,0.31 (1)  &1.98\,$\pm$\,0.24 (1)  &3.09\,$\pm$\,0.18 (2)  \\
\bottomrule
\end{tabular*}
\end{table*}

\setcounter{table}{2}

\begin{table*}
\centering
\caption{Continuation.}
\begin{tabular*}{0.9\linewidth}{@{\extracolsep{\fill}}lllllll}
\toprule
  Atomic &Elements& KIC\,9052363         & KIC\,9775385          & KIC\,10281360         & KIC\,11197934         &KIC\,11199412		\\ 
  number    &        & 			  &                       &                       &                       &    \\
\midrule
6&C   &8.80\,$\pm$\,0.14 (3) & 8.75\,$\pm$\,0.10 (6) &8.75\,$\pm$\,0.25 (4)  &8.95\,$\pm$\,0.37 (1)  &8.82\,$\pm$\,0.21 (2) \\
%$_{7}$N   &                      &		         &                       &		         &			    \\
8&O   &                      & 8.96\,$\pm$\,0.24 (1) &8.67\,$\pm$\,0.25 (1)  &			 &8.34\,$\pm$\,0.21 (1)  \\
11&Na &                      & 5.96\,$\pm$\,0.24 (1) &6.16\,$\pm$\,0.25 (1)	 &			 &  \\
12&Mg &7.56\,$\pm$\,0.10 (4) & 7.96\,$\pm$\,0.17 (5) &7.48\,$\pm$\,0.22 (5)  &8.09\,$\pm$\,0.37 (2)  &6.68\,$\pm$\,0.14 (3)  \\
14&Si &7.23\,$\pm$\,0.23 (6) & 7.16\,$\pm$\,0.38 (21)&7.01\,$\pm$\,0.10 (3)  &7.04\,$\pm$\,0.37 (2)  &6.07\,$\pm$\,0.21 (2) \\
16&S  &			 & 6.99\,$\pm$\,0.24 (2) &                       &                       &                      \\
20&Ca &6.20\,$\pm$\,0.22 (8) &6.37\,$\pm$\,0.29 (17) &6.35\,$\pm$\,0.19 (9)  &6.27\,$\pm$\,0.52 (5)  &5.86\,$\pm$\,0.24 (5) \\
21&Sc &2.98\,$\pm$\,0.15 (6) &2.94\,$\pm$\,0.22 (12) &2.96\,$\pm$\,0.12 (6)  &4.35\,$\pm$\,0.70 (3)  &2.18\,$\pm$\,0.21 (2)  \\
22&Ti &5.04\,$\pm$\,0.34 (23)&5.06\,$\pm$\,0.22 (40) &4.86\,$\pm$\,0.31 (29) &5.05\,$\pm$\,0.23 (8)  &4.64\,$\pm$\,0.35 (19) \\
23&V  &4.36\,$\pm$\,0.22 (2) &4.71\,$\pm$\,0.19 (6)  &4.03\,$\pm$\,0.28 (4)  &4.35\,$\pm$\,0.37 (1)  &3.67\,$\pm$\,0.12 (3)\\
24&Cr &5.64\,$\pm$\,0.26 (13)&5.75\,$\pm$\,0.19 (26) &5.46\,$\pm$\,0.29 (15) &5.75\,$\pm$\,0.18 (8)  &5.03\,$\pm$\,0.21 (7) \\
25&Mn &5.45\,$\pm$\,0.13 (3) &5.66\,$\pm$\,0.30 (11) &5.16\,$\pm$\,0.25 (2)  &5.38\,$\pm$\,0.37 (2)  &5.03\,$\pm$\,0.13 (4)  \\
26&Fe &7.40\,$\pm$\,0.19 (42)&7.44\,$\pm$\,0.17 (90) &7.29\,$\pm$\,0.14 (48) &7.46\,$\pm$\,0.21 (15) &6.68\,$\pm$\,0.24 (36)\\
%$_{27}$Co &                      &		         &                       &		         &                       \\
28&Ni &6.35\,$\pm$\,0.17 (6) &6.33\,$\pm$\,0.36 (27) &6.12\,$\pm$\,0.14 (13) &6.35\,$\pm$\,0.37 (1)  &5.73\,$\pm$\,0.24 (3)\\
29&Cu &                      &3.50\,$\pm$\,0.24 (2)  &                       &		         & \\
30&Zn &                      &			 &		         &			 &                      \\
38&Sr &2.11\,$\pm$\,0.22 (1) &3.13\,$\pm$\,0.24 (2)  &			 &			 & \\
39&Y  &2.65\,$\pm$\,0.22 (2) &2.51\,$\pm$\,0.31 (5)  &2.56\,$\pm$\,0.63 (4)	 &			 &2.36\,$\pm$\,0.21 (2) \\
40&Zr &3.14\,$\pm$\,0.22 (2) &3.30\,$\pm$\,0.10 (5)  &2.45\,$\pm$\,0.25 (2)  &			 &\\
56&Ba &2.09\,$\pm$\,0.22 (2) &1.99\,$\pm$\,0.17 (3)  &2.27\,$\pm$\,0.25 (2)  &3.66\,$\pm$\,0.37 (2)  &1.47\,$\pm$\,0.21 (2)  \\
\bottomrule
\end{tabular*}
\end{table*}

\setcounter{table}{2}

\begin{table*}
\centering
\caption{Continuation.}
\begin{tabular*}{0.9\linewidth}{@{\extracolsep{\fill}}llllll}
\toprule
  Atomic    &Elements& KIC\,11508397         & KIC\,11612274          & KIC\,11718839         & KIC\,11822666 \\
  number    &        & 			     &                        &                       &                         \\

\midrule
6&C   &8.89\,$\pm$\,0.31 (1) & 8.30\,$\pm$\,0.22 (6) &8.64\,$\pm$\,0.15 (6)  &8.95\,$\pm$\,0.37 (1) \\
%$_{7}$N   &                      &		         &                       &		         \\
8&O   &                      & 			 &8.82\,$\pm$\,0.25 (1)  &			\\
11&Na &                      & 			 &			 &			 \\
12&Mg &8.17\,$\pm$\,0.21 (4) & 7.86\,$\pm$\,0.10 (4) &7.93\,$\pm$\,0.18 (5)  &8.09\,$\pm$\,0.37 (2)  \\
14&Si &6.72\,$\pm$\,0.31 (2) & 7.41\,$\pm$\,0.10 (3) &7.25\,$\pm$\,0.40 (10) &7.04\,$\pm$\,0.37 (2)  \\
16&S  &			 &			 &                       &                       \\
20&Ca &7.73\,$\pm$\,0.41 (4) &6.57\,$\pm$\,0.39 (9)  &6.45\,$\pm$\,0.15 (17) &6.27\,$\pm$\,0.52 (5)  \\
21&Sc &			 &3.33\,$\pm$\,0.29 (5)  &3.31\,$\pm$\,0.35 (14) &4.35\,$\pm$\,0.70 (3)  \\
22&Ti &5.21\,$\pm$\,0.34 (7) &4.96\,$\pm$\,0.28 (19) &5.17\,$\pm$\,0.15 (40) &5.05\,$\pm$\,0.23 (8)  \\
23&V  &4.13\,$\pm$\,0.31 (1) &4.04\,$\pm$\,0.25 (2)  &4.40\,$\pm$\,0.25 (2)  &4.35\,$\pm$\,0.37 (1) \\
24&Cr &5.69\,$\pm$\,0.36 (5) &5.50\,$\pm$\,0.28 (12) &5.75\,$\pm$\,0.17 (24) &5.75\,$\pm$\,0.18 (8)  \\
25&Mn &5.07\,$\pm$\,0.31 (1) &5.34\,$\pm$\,0.25 (2)  &5.64\,$\pm$\,0.14 (5)  &5.38\,$\pm$\,0.37 (2)  \\
26&Fe &7.68\,$\pm$\,0.19 (23)&7.38\,$\pm$\,0.14 (36) &7.45\,$\pm$\,0.15 (106)&7.46\,$\pm$\,0.21 (15) \\
%$_{27}$Co &                      &		         &                       &		         \\
28&Ni &6.31\,$\pm$\,0.31 (2) &6.09\,$\pm$\,0.34 (10) &6.21\,$\pm$\,0.14 (15) &6.35\,$\pm$\,0.37 (1)  \\
29&Cu &                      &			 &                       &		          \\
30&Zn &                      &			 &		         &			 \\
38&Sr &1.65\,$\pm$\,0.31 (1) &3.09\,$\pm$\,0.24 (1)  &2.53\,$\pm$\,0.12 (3)	 &			  \\
39&Y  &			 &2.23\,$\pm$\,0.31 (2)  &2.21\,$\pm$\,0.19 (3)	 &			 \\
40&Zr &			 &3.36\,$\pm$\,0.10 (2)  &3.11\,$\pm$\,0.19 (4)  &			 \\
56&Ba &1.84\,$\pm$\,0.31 (1) &2.73\,$\pm$\,0.17 (2)  &2.24\,$\pm$\,0.10 (3)  &3.66\,$\pm$\,0.37 (2)  \\
\bottomrule
\end{tabular*}
\end{table*}

\end{document}